\documentclass[Conference,a4paper]{IEEEtran}
\IEEEoverridecommandlockouts
\usepackage{color}
\usepackage{graphicx}
\usepackage{epstopdf}
\usepackage{amsmath}
\usepackage{amssymb}
\usepackage{algorithm}
\usepackage{algorithmic}
\usepackage{amsmath}
\usepackage{multirow}
\usepackage{booktabs}
\usepackage{array}
\usepackage{amsthm}
\usepackage{stfloats}
\usepackage{caption}
\usepackage{subfigure}
\usepackage{bm}
\usepackage{booktabs}
\usepackage{setspace}
\usepackage{diagbox}
\usepackage{enumerate}
\usepackage{ulem}
\usepackage{url}

\allowdisplaybreaks[4]

{\bgroup
 \addtolength\abovedisplayshortskip{#1}
 \addtolength\abovedisplayskip{#1}
 \addtolength\belowdisplayshortskip{#1}
 \addtolength\belowdisplayskip{#1}
 }
{\egroup\ignorespacesafterend}

\newtheorem{theorem}{Theorem}
\newtheorem{lemma}{Lemma}

\newcommand{\be}{\begin{equation}}
\newcommand{\ee}{\end{equation}}
\newcommand{\bea}{\begin{eqnarray}}
\newcommand{\eea}{\end{eqnarray}}
\newcommand{\ba}{\begin{array}}
\newcommand{\ea}{\end{array}}

\title{Joint Beamforming and Power Allocation for RIS Aided Full-Duplex Integrated Sensing and Uplink Communication System}
\author{\IEEEauthorblockN{Yuan Guo, Yang Liu, Qingqing Wu, Xiaoyang Li,  and Qingjiang Shi
\thanks{
Y. Guo and Y. Liu  are with the School of Information
and Communication Engineering, Dalian University of Technology, Dalian,
China, email: yuanguo@mail.dlut.edu.cn,
yangliu\_613@dlut.edu.cn.
}
\thanks{
Q. Wu is with Department of Electronic Engineering, Shanghai Jiao Tong University, Shanghai, China, email: qingqingwu@sjtu.edu.cn.
}
\thanks{
X. Li is  with the Shenzhen Research Institute of Big Data,
Shenzhen, China, email: lixiaoyang@sribd.cn.
}
\thanks{
Q. Shi is with the School of Software Engineering, Tongji University,
Shanghai, China, and also with the Shenzhen Research Institute of Big Data,
Shenzhen, China, email: shiqj@tongji.edu.cn.
}
}
} 

\begin{document}
\maketitle
\pagestyle{empty}
\thispagestyle{empty}

\begin{abstract}
Integrated sensing and communication (ISAC) capability is envisioned as one key feature for future cellular networks.
Classical half-duplex (HD) radar sensing is conducted in a ``first-emit-then-listen'' manner.
One challenge to realize HD ISAC lies in the discrepancy of the two systems' time scheduling for transmitting and receiving.
This difficulty can be overcome by full-duplex (FD) transceivers.
Besides,
ISAC generally has to comprise its communication rate due to realizing sensing functionality.
This loss can be compensated by the emerging reconfigurable intelligent surface (RIS) technology.
This paper considers the joint design of beamforming, power allocation and signal processing in a FD uplink communication system aided by RIS, which is a highly nonconvex problem.
To resolve this challenge,
via leveraging the cutting-the-edge majorization-minimization (MM) and penalty-dual-decomposition (PDD) methods,
we develop an iterative solution that optimizes all variables via using convex optimization techniques.
Besides,
by wisely exploiting alternative direction method of multipliers (ADMM) and optimality analysis,
we further develop a low complexity solution that updates all variables analytically and runs highly efficiently.
Numerical results are provided to verify the effectiveness and efficiency of our proposed algorithms and demonstrate the significant performance boosting by employing RIS in the FD ISAC system.

\end{abstract}


\maketitle
\section{Introduction}
Recently,
the integrated sensing and communication (ISAC) system has attracted great attentions from both industry and academia \cite{ref1}$-$\cite{ref3}.
On the one hand,
next generation cellular system featured by millimeter wave (mmWave) and Terahertz communication techniques will occupy wide high-frequency bands,
which overlap with those for radar systems.
On the other hand,
the booming Internet of Things (IoT) applications require mobile devices
to become more functional and possess sensing capability.
In this context,
ISAC has been envisioned as a promising solution,
which aims at realizing both sensing and communication functionalities using one unified hardware set and sharing frequency spectrums.
Many latest progresses for joint radar sensing and communication design can be found in \cite{ref1}$-$\cite{ref3} and the reference therein.

Despite its hardware and spectral efficiency,
the ISAC system's dual functionalities generally come at a cost of compromising performance in both sensing the communication.
This drawback can hopefully be overcome by the emerging reconfigurable intelligent surface (RIS) technology \cite{ref8},
which is also widely known as intelligent reflecting surface (IRS) \cite{ref10}.
The RIS is envisioned as a viable approach to enhance communication system.
It can empower the wireless system with additional beamforming capability via reflecting and adjusting phase shifts of the incoming signals at a relatively low energy and hardware expense.
The versatility of RIS in boosting communication performance in various aspects have been extensively verified recently, see \cite{ref8} and \cite{ref10}
and their reference.

\subsection{Related Works}
Due to the aforementioned advantage of the ISAC system and the merits of RIS technology,
a rich body of literature has studied deploying the RIS in ISAC context
and conducted joint design to improve the sensing and communication performance,
e.g., \cite{ref11}$-$\cite{ref29.1.3}.
For instance, the authors of \cite{ref11} and \cite{ref12} first proposed to leverage RIS in ISAC system and demonstrated that RIS could effectively reduce multi-user interference (MUI)  and Cram\'{a}r-Rao bound (CRB) of the radar sensing signal, respectively.
The work \cite{ref13} showed that, via deploying RIS, joint active and passive beamforming design could significantly improve the signal-to-noise-ratio (SNR) of radar signal processing while guaranteeing the quality-of-service (QoS) of mobile users.
The authors of
\cite{ref15}
proposed a RIS-aided waveform design towards maximizing the weighted sum of radar sensing SNR and information receiving SNR,
which was shown to boost both functionals.
The works \cite{ref16}  utilized RIS to improve communication performance under the cross-correlation constraint.
The authors of \cite{ref19} considered sensing target with non-negligible shape and illustrated that RIS could greatly enhance the ultimate detection resolution (UDR) of target detection.
{
The authors of \cite{ref20.1} developed a low-complexity RIS beamforming algorithm which could maximize both the communication and sensing SNR.}
The paper \cite{ref20.2} showed that RIS could effectively elevate the radar mutual information (MI), which is generally a good performance metric for both detection and estimation.
The recent works \cite{ref21.1}$-$\cite{ref22} employed the novel simultaneously transmitting and reflecting (STAR) architecture to effectively extend the RIS' coverage in ISAC network.
Lately, the authors of \cite{ref22.2} adopted the emerging active RIS architecture in ISAC system and showed it could significantly boost signal-to-interference-plus-noise-ratio (SINR) of radar sensing compared to the passive RIS.
The papers \cite{ref24}$-$\cite{ref25} demonstrated that RIS could remarkably enhance the security of sensing when the probing signal contained communication information.
{The recent work \cite{ref26} proposed  four different algorithms to enhance the RIS-aided ISAC system's performance.}
The works \cite{ref27} and \cite{ref29} demonstrated that the deployment of RIS could remarkably inflate the beampattern gain in non-orthogonal multiple access (NOMA) and millimeter wave ISAC networks, respectively.
Besides,
very recently,
several latest works paid attention to implementing ISAC systems utilizing full-duplex (FD) BSs
\cite{ref29.1.1}$-$\cite{ref29.1.3}.
The authors of \cite{ref29.1.1} optimized FD BS' hybrid precoder to enhance mobile users' spectral efficiency and radar sensing capability.
The work \cite{ref29.1.2} designed FD ISAC system's waveform to improve radar detection probability via suppressing self-interference and enhancing autocorrelation.
In \cite{ref29.1.3},
the authors developed secure communication beamforming aided by RIS in a uplink communication system with FD BS

\subsection{Motivations and Contributions}
As seen above,
although a rich body of existing literature has investigated the waveform design for ISAC system,
most of these works have considered half-duplex (HD) systems,
 where transmitting and receiving (T\&R) are operated separately in time.
 Note the switching frequencies between T\&R are usually different for communication and radar sensing,
 which is indeed one challenge to implement ISAC system \cite{ref3}.
 For instance, the 3GPP NR specification \cite{ref29_3} has supported flexible UL/DL mini-slot frame structures,
 which will inevitably lead to more unpredictable T\&R switching frequency for communication.
 In contrast,
 the FD system tends to be a highly promising solution to accommodate the aforementioned conflict \cite{ref2}$-$\cite{ref3}.
 Note the waveform design in FD ISAC system has far from being thoroughly investigated,
 except the small number of latest works \cite{ref29.1.1}$-$\cite{ref29.1.3}.
 More importantly,
 all the existing works \cite{ref29.1.1}$-$\cite{ref29.1.3} have not considered the deployment of RIS in the FD system.
 Based on the above inspections, we are motivated to study a RIS-aided FD ISAC system to fully promote its communication and sensing capabilities.
 Specifically, the contributions of this paper are elaborated as follows:
%
%
%

\begin{itemize}
\item
This paper considers the joint beamforming design in a FD ISAC system aided by RIS to realize simultaneous UL communication and target sensing. We study maximizing the sum-rate of all UL users while assuring the radar sensing quality via designing BS probing beamforming, RIS phase-shifts, users power allocation and receiving processors. To the best of our knowledge, this problem has not been considered in the existing literature, e.g., \cite{ref11}$-$\cite{ref29.1.3}.

\item
Moreover,
this paper considers a very generic signal propagation model,
which fully takes into account the RIS effect in both the forwarding and reflected radar probing signals.
As will be seen,
this consideration in modeling significantly complicates the beamforming design task and yields a highly challenging quartic fractional programming problem.

\item
To attack the above challenge,
we develop an alternative optimization algorithm that optimizes all variables via convex optimization techniques.
Especially,
to tackle the highly challenging RIS configuration problem,
by wisely introducing splitting variable and leveraging the penalty dual decomposition (PDD) \cite{ref33} framework combined with the majorization-minimization (MM) \cite{ref32} method,
we obtain a solution to resolve the quartic optimization by solving a series of quadratic sub-problems.
This method is never seen in the existing literature.

\item
Furthermore,
we also develop low complexity solution.
By exploiting alternative direction method of multipliers (ADMM) \cite{ref35} and analyzing optimality conditions,
we succeed in optimizing all blocks of variables analytically.
Our proposed analytic-updated solution does not resort to any numerical solver,
e.g., CVX \cite{ref34}, and is rarely seen in the existing literature, e.g., \cite{ref11}$-$\cite{ref29.1.3}.

\item
Last but not least,
extensive numerical results are provided to verify the effectiveness and efficiency of our proposed solutions.
At the same time,
experiment results demonstrate that the deployment of RIS can significantly benefit the UL communication in the considered FD ISAC scenario.

\end{itemize}

The rest of the  paper is organized as follows.
Section \uppercase\expandafter{\romannumeral2} will introduce the model of a FD ISAC system assisted by RIS and formulate the joint  beamforming design problem.
Section \uppercase\expandafter{\romannumeral3}  will propose an iterative solution to tackle the proposed beamforming design problem.
A low complexity algorithm will be developed in Section \uppercase\expandafter{\romannumeral4}.
Section \uppercase\expandafter{\romannumeral5} and
Section \uppercase\expandafter{\romannumeral6} will  present numerical  results and conclude the paper, respectively.

\section{System Model and Problem Formulation}
\subsection{System Model}
\begin{figure}[t]
	\centering
	\includegraphics[width=.4\textwidth]{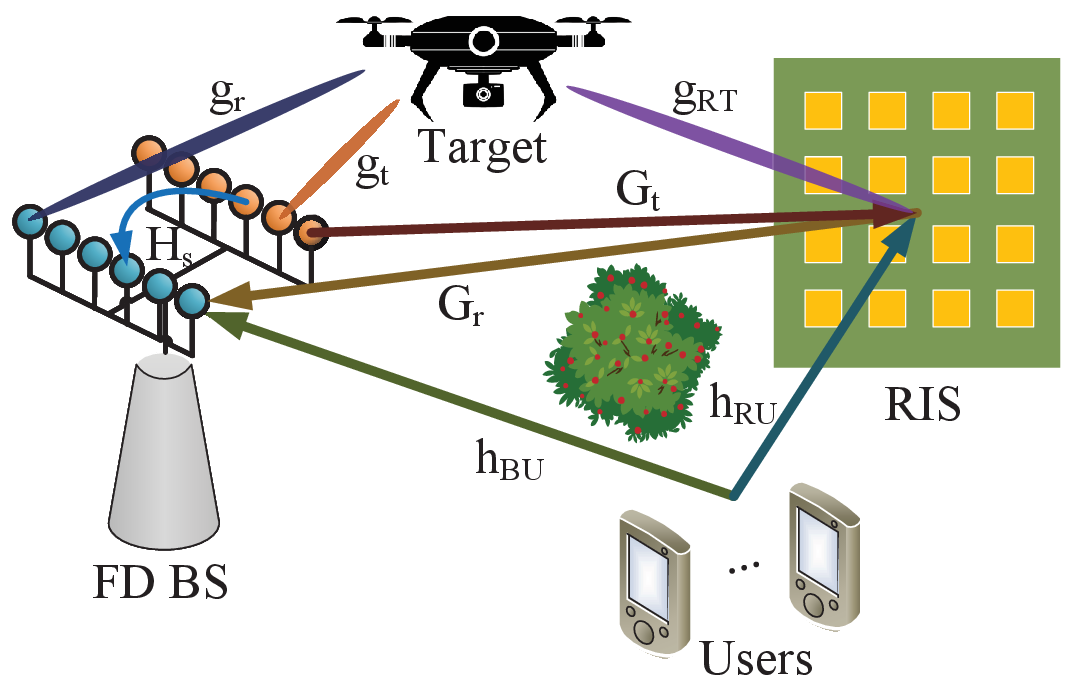}
	\caption{An  RIS-aided  FD ISAC  system.}
	\label{fig.1}
\end{figure}
{As shown in Fig. \ref{fig.1},
we consider an uplink multi-user MISO RIS-aided ISAC system
consisting a FD BS equipped with $N_t$ transmit (TX) antennas and $N_r$ receiver (RX) antennas,}
an RIS  with $M$ reflecting units, $K$ single-antenna uplink mobile users and one point-like target\footnote{
In fact, the solution developed in this paper can be easily extended to the multi-target scheme.
Due to space of limit, we leave the multi-target case for future study.}.
The BS with aid of an RIS simultaneously receives the information from  mobile users  and transmits  probing waveform to detect the target.
For convenience,
the sets of users and RIS units are denoted by $\mathcal{K}=\{1,\cdots,K\}$ and $\mathcal{M}=\{1,\cdots,M\}$, respectively.
Besides,
we assume that the transmission links between target and users are blocked.

The uplink signal  transmitted by the $k$-th uplink mobile user can be  written as
\begin{align}
x_{u,k}=\sqrt{q_k}s_{u,k}, \forall k \in \mathcal{K},
\end{align}
where $\mathcal{K}\triangleq\{1,\cdots,K\}$,
$s_{u,k}$ and $q_k$ are the information symbol and transmission power of $k$-th user, respectively.
For simplicity,
we assume that $s_{u,k}$ are  mutually uncorrelated and each has zero mean and unit variance.

To conduct target sensing,
the BS transmits probing signal,
which is given as \cite{ref_II_A_1}-\cite{ref_II_A_2}
\begin{align}
\mathbf{x}=\mathbf{W}\mathbf{s}_r,
\end{align}
where
the vector
$\mathbf{s}_r \in \mathbb{C}^{N_t\times 1}$
denotes  radar probing signal
and
has zero mean and covariance matrix $\mathbb{E}\{\mathbf{s}_r\mathbf{s}_r^H\}=\mathbf{I}_{N_t}$,
and $\mathbf{W}\in \mathbb{C}^{N_t\times N_t}$ represents the beamformer for the probing signal.

{As shown in Fig. \ref{fig.1},
we denote the wireless links of
BS TX-RIS,
BS RX-RIS,
BS TX-user-$k$,
RIS-user-$k$,
RIS-target
and
the self-interference (SI) of the BS as
$\mathbf{G}_t \in \mathbb{C}^{M \times N_t}$,
$\mathbf{G}_r \in \mathbb{C}^{M \times N_r}$,
$\mathbf{h}_{BU,k} \in \mathbb{C}^{N_r \times 1}$,
$\mathbf{h}_{RU,k} \in \mathbb{C}^{M \times 1}$,
$\mathbf{g}_{RT}\in \mathbb{C}^{M \times 1}$
and
$\mathbf{H}_s \in \mathbb{C}^{N_t \times N_r}$, respectively.}
The phase-shifting conducted by the RIS elements to their impinging signals can be modeled as a complex vector
${\bm{\phi}}=[e^{j\theta_1},\dots,e^{j\theta_M}]^T$,
with $\theta_m$ representing the phase shift of $m$-th reflecting unit,
$\theta_m \in [0, 2\pi)$ and $\forall m \in \mathcal{M}$.
In the following, we will alternatively use the diagonal matrix
$\bm{\Phi}\triangleq\mathrm{diag}(\bm{\phi})$ to represent the RIS' reflection coefficients.

The  steering vectors of BS can be expressed as
\begin{small}
\begin{subequations}
\begin{align}
&\!\!\bm{a}_t\!\! =\! [1,e^{-j2\pi d \sin(\theta_t)/\lambda},\cdots,e^{-j2\pi d(N_t\!-\!1)\sin(\theta_t)/\lambda}]^T\!\in\!\mathbb{C}^{N_t\! \times\! 1},\\
&\!\!\bm{a}_r\!\! =\! [1,e^{-j2\pi d \sin(\theta_r)/\lambda},\cdots,e^{-j2\pi d(N_r\!-\!1)\sin(\theta_r)/\lambda}]^T\!\!\!\in\!\mathbb{C}^{N_r\! \times\! 1},
\end{align}
\end{subequations}
\end{small}
\vspace{-0.5cm}

\noindent
respectively,
where $d$ denoting the antenna spacing and $\lambda$ denoting the carrier wavelength.
$\theta_{t} $ and $\theta_{r}$ are the angle of departure (AoD)  and angle of arrival (AoA) with respect to transmit and receive antennas of the BS, respectively.

{
The steering vector of RIS is expressed as
\begin{align}
&\bm{a}_{RT}(\bm{g}_{R}) = \bm{a}_{M_1}({g}_{R,1}) \otimes \bm{a}_{M_2}({g}_{R,2}),
\end{align}
with
\begin{small}
\begin{subequations}
\begin{align}
&\!\!\bm{g}_{R}\!\! =\!\! \big\{{g}_{R,1}\!\!=\!\! \frac{1}{2}\textrm{sin}(\theta_{R,e})\textrm{cos}(\theta_{R,z}),
{g}_{R,2}\!\!=\!\! \frac{1}{2}\textrm{cos}(\theta_{R,e}) \!  \big\},\\
&\!\!\bm{a}_{\!M_1}\!(\!{g}_{R,1}\!)\!\!=\!\! [1,e^{j2\pi d {g}_{R,1}/ \lambda},\cdots,e^{j2\pi d(M_1\!-\!1){g}_{RT,1}/ \lambda}]^T\!\!\!\in\!\!\mathbb{C}^{M_1\! \times\! 1},\\
&\!\!\bm{a}_{\!M_2}\!(\!{g}_{R,2}\!)\!\!=\!\! [1,e^{j2\pi d {g}_{R,2}/ \lambda},\cdots,e^{j2\pi d(M_2\!-\!1){g}_{RT,2}/ \lambda}]^T\!\!\!\in\!\!\mathbb{C}^{M_2\! \times\! 1},
\end{align}
\end{subequations}
\end{small}}
\vspace{-0.3cm}

\noindent
{where $\theta_{R,e} $ and $\theta_{R,z} $ represent  elevation and azimuth angles of the angle of departure (AoD) of the RIS,  respectively.
Besides,
$M_1$ and
$M_2$ denote the numbers of elevation and azimuth RIS' elements,
respectively,
and the total number of RIS' elements is $M=M_1\times M_2$.}

{Therefore, the BS TX-target, the BS RX-target and RIS-target channels are given as
\begin{small}
\begin{align}
\mathbf{g}_t\!\!=\!\!\alpha_t\bm{a}_t\!\in\! \mathbb{C}^{N_t \!\times\! 1},
\mathbf{g}_r\!\!=\!\!\alpha_r\bm{a}_r\!\in\! \mathbb{C}^{N_r \!\times\! 1},
\mathbf{g}_{RT}\!\!=\!\!\alpha_{RT}\bm{a}_{RT}\!\in\! \mathbb{C}^{M \!\times\! 1},
\end{align}
\end{small}}
\vspace{-0.5cm}

\noindent
{respectively,
where $\alpha_{t}$, $\alpha_{r}$ and $\alpha_{RT}$  are  complex fading coefficients and  are assumed to be known.}

To perform target sensing,
the BS emits probing waveform towards the target and  listens to its echoes rebounding from the target simultaneously.
During the whole procedure,
all mobile users are operating in uplink mode and transmitting information symbols to the BS.
The received signal at FD BS can be represented by
\begin{small}
\begin{align}
\mathbf{y}=
&\underbrace{{\sum}_{k=1}^{K}(\mathbf{h}_{BU,k}\!+\!\mathbf{G}_r^H\bm{\Phi}\mathbf{h}_{RU,k})x_{u,k}}
\limits_{\textrm{UL}\ \textrm{communication}\ \textrm{signal}}
+\underbrace{(\mathbf{G}_r^H\bm{\Phi}\mathbf{G}_t+\mathbf{H}_{s}^H)\mathbf{x}}\limits_{\textrm{self-interference}} \label{received signal}\\ \nonumber
&+\underbrace{\alpha(\mathbf{g}_r+\mathbf{G}_r^H\bm{\Phi}\mathbf{g}_{RT})(\mathbf{g}_t^H+\mathbf{g}_{RT}^T\bm{\Phi}\mathbf{G}_t)\mathbf{x}}\limits_{\textrm{sensing echoes}}
+\mathbf{n}_{BS},
\end{align}
\end{small}
\vspace{-0.4cm}

\noindent
where
$\alpha$ denotes the target radar cross section (RCS) and $\mathbb{E}\{\vert \alpha\vert^2\}= \sigma_{t}^2$,
$\mathbf{n}_{BS} \sim \mathcal{CN}(0,\mathbf{I}_{N_r}) $ is the complex additive white Gaussian noise (AWGN) at the BS.
Note in (\ref{received signal}),
the signals reflected more than thrice are neglected due to the severe attenuations.

To recover different users' information and improve target sensing performance,
the BS utilizes $K+1$ linear filters
$\mathbf{u}_j \in \mathbb{C}^{N_r \times 1}$,
$ \forall j \in \mathcal{{J}} \triangleq \{0\}\bigcup\mathcal{K}$
to post-process the received signal,
where index $0$ corresponds to the radar sensing filter bank.
Therefore,
the output of the $j$-th post-processor is given as
\begin{align}
{y}_j = \mathbf{u}_j^H\mathbf{y}, \forall j \in \mathcal{{J}}.
\end{align}

The SINR for the $k$-th mobile user can be readily obtained as
\begin{align}
&{\textrm{SINR}_{U,k}(\mathbf{W},\mathbf{u}_k,\{q_k\},\bm{\phi})=}\\
&{\frac{q_k\vert\mathbf{u}^H_k\mathbf{h}_{U,k}\vert^2}
{\sum_{i\neq k}^{K}\! q_i\!\vert\!\mathbf{u}^H_k\mathbf{h}_{U,i}\!\vert^2
\!\!+\!\!\sigma_t^2\Vert\!\mathbf{u}^H_k\mathbf{H}(\bm{\Phi})\mathbf{W}\!\Vert^2_2
\!\!+\!\!\Vert\!\mathbf{u}^H_k\mathbf{G}\mathbf{W}\!\Vert^2_2
\!\!+\!\!\sigma^2_r\Vert\!\mathbf{u}_k^H\!\Vert^2_2
},}\nonumber
\end{align}
where $\mathbf{h}_{U,k} \triangleq \mathbf{h}_{BU,k}+\mathbf{G}^H_r\bm{\Phi}\mathbf{h}_{RU,k}$,
$\mathbf{G}\triangleq\mathbf{G}_r^H\bm{\Phi}\mathbf{G}_t+\mathbf{H}_{s}^H$ and
$\mathbf{H}(\bm{\Phi})\triangleq (\mathbf{g}_r+\mathbf{G}_r^H\bm{\Phi}\mathbf{g}_{RT})(\mathbf{g}_t^H+\mathbf{g}_{RT}^T\bm{\Phi}\mathbf{G}_t)   $.

The achievable rate of  each user is given as
\begin{align}
\textrm{R}_k(\mathbf{W},\mathbf{u}_k,\{q_k\},\bm{\phi})
 = \textrm{log}(1+\textrm{SINR}_{U,k}), \forall k \in \mathcal{K}.
\end{align}

The output SINR for target sensing can  be given by
\begin{align}
&\textrm{SINR}_{r}(\mathbf{W},\mathbf{u}_0,\{q_k\},\bm{\phi})=\\
&\frac{\sigma_t^2\Vert\mathbf{u}^H_0{\mathbf{H}(\bm{\Phi})}\mathbf{W}\Vert^2_2
}
{\sum_{k=1}^{K} q_k\vert\mathbf{u}^H_0\mathbf{h}_{U,k}\vert^2
+\Vert\mathbf{u}^H_0\mathbf{G}\mathbf{W}\Vert^2_2
+\sigma^2_r\Vert\mathbf{u}_0^H\Vert^2_2}.\nonumber
\end{align}

\subsection{Problem Formulation}
Our goal is to maximize the sum-rate of all users via jointly optimizing
the transmit beamformer $\mathbf{W}$,
the linear post-processing filters ($\{\mathbf{u}_k\}$, $\mathbf{u}_0$),
the users' uplink transmit power $\{q_k\}$
and the reflected phase shift  $\bm{\phi}$.
The optimization problem can be formulated as
\begin{subequations}
\begin{align}
\textrm{(P0)}:\mathop{\textrm{max}}
\limits_{\mathbf{W},
\{\mathbf{u}_k\},
\mathbf{u}_0,
\{q_k\},
\bm{\phi}}\
&{\sum}_{k=1}^{K} \textrm{R}_k(\mathbf{W},\mathbf{u}_k,\{q_k\},\bm{\phi})\label{P0_obj}\\
\textrm{s.t.}\
&\textrm{SINR}_{r}(\mathbf{W},\mathbf{u}_0,\{q_k\},\bm{\phi})\!\geq \!\Gamma_r,\\
&\Vert\mathbf{W}\Vert_F^2\leq P_{BS},\\
&0 \leq q_k\leq P_{U,k}, \forall k \in \mathcal{K},\\
&\vert\phi_m\vert=1, \forall m \in \mathcal{M},
\end{align}
\end{subequations}
where $\Gamma_r$ and $P_{BS}$
denote the predefined target sensing performance threshold
and the maximum transmission power  of the BS, respectively,
and
$P_{U,k}$ is the  uplink communication power budget of $k$-th user.
The problem (P0) is highly challenging due to its highly non-convex
objective and constraints.

\section{SOCP-based Algorithm}
\subsection{Problem Reformulation}
In order to make the problem (P0) more tractable,
we firstly employ the weighted minimum mean squared error (WMMSE) method \cite{ref31} to transform its objective function.
Specifically,
via introducing auxiliary variables $\{\beta_k\}$ and $\{\omega_k\}$,
the original objective function (\ref{P0_obj}) can be equivalently written into a variation form
(\ref{MSE}) \cite{ref31}, as shown on the top of next page.
\begin{figure*}
\begin{small}
\begin{align}
&{\textrm{R}_k(\mathbf{W},\mathbf{u}_k, \{q_k\},\bm{\phi})}\label{MSE}\\
&{=\mathop{\textrm{max}}
\limits_{
\omega_k\geq0
}
\textrm{log}(\omega_k)-\omega_k\big({\sum}_{i=1}^{K}q_i\vert\mathbf{u}^H_k\mathbf{h}_{U,k}\vert^2
+\sigma_t^2\Vert\mathbf{u}^H_k\mathbf{H}(\bm{\Phi})\mathbf{W}\Vert^2_2
+\Vert\mathbf{u}^H_k\mathbf{G}\mathbf{W}\Vert^2_2
+\sigma_r^2\Vert\mathbf{u}^H_k\Vert^2_2\big)^{-1}\sqrt{q_k}\mathbf{u}^H_k\mathbf{h}_{U,k}+1}\nonumber\\
&{=\!\!\mathop{\textrm{max}}
\limits_{
\omega_k\geq0,
\beta_k
}
\!\underbrace{\textrm{log}(\omega_k)
\!\!-\!\!\omega_k\bigg(1\!\!-\!\!2\textrm{Re}\{\beta_k^{\ast}\sqrt{q_k}\mathbf{u}^H_k\mathbf{h}_{U,k}\}
\!\!+\!\!\vert\beta_k\vert^2\big({\sum}_{i=1}^{K}q_i\vert\mathbf{u}^H_k\mathbf{h}_{U,k}\vert^2
\!\!+\!\!\sigma_t^2\Vert\mathbf{u}^H_k\mathbf{H}(\bm{\Phi})\mathbf{W}\Vert^2_2
\!\!+\!\Vert\mathbf{u}^H_k\mathbf{G}\mathbf{W}\Vert^2_2
\!\!+\!\sigma_r^2\Vert\mathbf{u}^H_k\!\Vert^2_2\big)\bigg)\!\!+\!\!1}\limits_{\mathrm{\tilde{R}}_k(\mathbf{W},\mathbf{u}_k,\{q_k\},\bm{\phi},\omega_k,\beta_k)}, \forall k\!\! \in\!\! \mathcal{K}.}\nonumber
\end{align}
\end{small}
\boldsymbol{\hrule}
\end{figure*}
Therefore,
the original problem (P0) is equivalently expressed as
\begin{subequations}
\begin{align}
\!\!\!\!\!\textrm{(P1)}:\!\!\!\!\!\!\!\!\mathop{\textrm{max}}
\limits_{\mathbf{W},
\{\!\mathbf{u}_k\!\},
\mathbf{u}_0,
\atop
\{\!q_k\!\},
\bm{\phi},
\{\!\omega_k\!\},
\{\!\beta_k\!\}
}\
&\!\!\!\!{\sum}_{k\!=\!1}^{K}\! \mathrm{\tilde{R}}_k(\mathbf{W},\{\!\mathbf{u}_k\!\},\{\!q_k\!\},\bm{\phi},\omega_k,\beta_k)\label{P1_obj}\\
\textrm{s.t.}\
&\textrm{SINR}_{r}(\mathbf{W},\mathbf{u}_0,\{q_k\},\bm{\phi})\geq \Gamma_r,\label{P1_SINR_radar}\\
&\Vert\mathbf{W}\Vert_F^2\leq P_{BS},\label{P1_power_BS}\\
& 0 \leq q_k\leq P_{U,k}, \forall k \in \mathcal{K},\label{P1_power_user}\\
&\vert\phi_m\vert=1, \forall m \in \mathcal{M}\label{P1_phi_module}.
\end{align}
\end{subequations}

In the next, we adopt the block coordinate ascent (BCA) \cite{ref351_BCA} method to tackle the problem (P1).

\subsection{Optimizing auxiliary variables}
According to the derivation of WMMSE transformation,
with other variables being fixed,
the update of the auxiliary variables $\{\beta_k\}$ and $\{\omega_k\}$ have analytical solutions that are given as follows
\begin{small}
\begin{align}
&\!\!\!\!\beta_k^{\star}\!\!=\!\!\frac{\sqrt{q_k}\mathbf{u}_k^H\mathbf{h}_{U,k}}
{{\sum}_{i=1}^{K}q_i\vert\mathbf{u}^H_k\mathbf{h}_{U,i}\vert^2
\!\!+\!\!\sigma_t^2\Vert\!\mathbf{u}^H_k\mathbf{H}\mathbf{W}\!\Vert^2_2
\!\!+\!\!\Vert\!\mathbf{u}^H_k\mathbf{G}\!\mathbf{W}\!\Vert^2_2
\!\!+\!\!\sigma_r^2\Vert\!\mathbf{u}^H_k\!\Vert^2_2},\label{Solution_beta}\\
&\!\!\!\!\omega_k^{\star}
\!\!=\!\!1\!\!+\!\!\frac{{q_k}\mathbf{h}_{U,k}^H\mathbf{u}_k\mathbf{u}_k^H\mathbf{h}_{U,k}}
{{\sum}_{i\neq k}^{K}\!q_i\!\vert\!\mathbf{u}^H_k\mathbf{h}_{U,i}\!\vert^2
\!\!+\!\!\sigma_t^2\Vert\!\mathbf{u}^H_k\!\mathbf{H}\!\mathbf{W}\!\Vert^2_2
\!\!+\!\!\Vert\!\mathbf{u}^H_k\!\mathbf{G}\!\mathbf{W}\!\Vert^2_2
\!\!+\!\!\sigma_r^2\Vert\!\mathbf{u}^H_k\!\Vert^2_2}\label{Solution_omega}.
\end{align}
\end{small}

\subsection{Optimizing The Phase Shift}
In this subsection,
we investigate the optimization of the RIS phase-shifting $\bm{\phi}$
when other variables are given.
By introducing the  new coefficients as follows
\begin{align}
&\mathbf{P}_k
\triangleq
\mathbf{G}_r^H\textrm{diag}(\mathbf{h}_{RU,k}),
\mathbf{r}_k
\triangleq
\mathbf{G}_r\mathbf{u}_k,
\mathbf{v}_k
\triangleq
\mathbf{u}_k^H\mathbf{H}_{s}^H\mathbf{W},\\
&\mathbf{S}_k
\triangleq
\mathbf{W}^H\mathbf{G}_t^H\textrm{diag}(\mathbf{r}_k),
\mathbf{r}_0
\triangleq
\mathbf{G}_r\mathbf{u}_0,
\mathbf{v}_0
\triangleq
\mathbf{u}_0^H\mathbf{H}_{s}^H\mathbf{W},
\nonumber\\
&\mathbf{S}_0
\!\!\triangleq\!\!
\mathbf{W}^H\mathbf{G}_t^H\textrm{diag}(\mathbf{r}_0),
\mathbf{a}_{1,k}
\!\triangleq\!
\mathbf{W}^H\mathbf{g}_t\mathbf{g}_r^H\mathbf{u}_k,
\mathbf{b}_{3}
\!\!\triangleq\!\!
\mathbf{W}^H\mathbf{g}_t,
\nonumber\\
&b_{1,k}\!
\!\triangleq\!\!
\mathbf{g}_r^H\!\mathbf{u}_k,
\mathbf{B}
\!\!\triangleq\!\!
\textrm{diag}(\mathbf{g}_{RT}^H)\mathbf{G}_t\mathbf{W},
\mathbf{b}_{2,k}
\!\!\triangleq\!\!
\textrm{diag}(\mathbf{g}_{RT}^H)\mathbf{G}_r\mathbf{u}_k,
\nonumber\\
&\mathbf{a}_{0}
\!\triangleq\!
\mathbf{W}^H\mathbf{g}_t\mathbf{g}_r^H\mathbf{u}_0,
b_{0,1}
\!\triangleq\!
\mathbf{g}_r^H\mathbf{u}_0,
\mathbf{b}_{0,2}
\!\!\triangleq\!\!
\textrm{diag}(\mathbf{g}_{RT}^H)\mathbf{G}_r\mathbf{u}_0,\nonumber
\end{align}
the objective function (\ref{P1_obj}) and the constraint (\ref{P1_SINR_radar}) are,
respectively,
rewritten in (\ref{Phi_obj_trans}) and (\ref{Phi_SINR_trans}) as follows
\begin{subequations}
\begin{align}
&-{\sum}_{k=1}^{K} \mathrm{\tilde{R}}_k
= \bm{\phi}^H\mathbf{T}_1\bm{\phi}
-2\textrm{Re}\{\mathbf{t}^H_1\bm{\phi}\}+c_1\label{Phi_obj_trans}\\
&\qquad\qquad\quad+2\textrm{Re}\{\bm{\phi}^H\mathbf{T}_{1,5}\bm{\phi}^{\ast}
+\textrm{vec}(\bm{\phi}\bm{\phi}^T)^H\mathbf{T}_{1,67}\bm{\phi}\}\nonumber\\
&\qquad\qquad\quad+\textrm{vec}(\bm{\phi}\bm{\phi}^T)^H\mathbf{T}_{1,8}\textrm{vec}(\bm{\phi}\bm{\phi}^T),\nonumber\\
&\bm{\phi}^H\mathbf{T}_0\bm{\phi}
-2\textrm{Re}\{\mathbf{t}_0^H\bm{\phi}\}+c_2
-\bm{\phi}^H\mathbf{T}_{0,0}\bm{\phi}\label{Phi_SINR_trans}\\
&-2\textrm{Re}\{\bm{\phi}^H\mathbf{T}_{0,5}\bm{\phi}^{\ast}
+\textrm{vec}(\bm{\phi}\bm{\phi}^T)^H\mathbf{T}_{0,67}\bm{\phi}\}\nonumber\\
&-\textrm{vec}(\bm{\phi}\bm{\phi}^T)^H\mathbf{T}_{0,8}\textrm{vec}(\bm{\phi}\bm{\phi}^T)
\leq 0,\nonumber
\end{align}
\end{subequations}
with the parameters in (\ref{Phi_obj_trans}) and (\ref{Phi_SINR_trans})
being defined in (\ref{Phi_trans_coefficient}), as shown on the top of next page.
\begin{figure*}
\begin{small}
\begin{align}
& c_{1,0}
\triangleq
-{\sum}_{k=1}^{K}
\{
\textrm{log}(\omega_k)-\omega_k+1
+2\textrm{Re}\{\omega_k\beta_k^{\ast}\sqrt{q_k}\mathbf{u}_k^H\mathbf{h}_{BU,k}\}
-\omega_k\vert\beta_k\vert^2
[
{\sum}_{i=1}^{K}q_i\mathbf{u}_k^H\mathbf{h}_{BU,i}\mathbf{h}_{BU,i}^H\mathbf{u}_k
+
\mathbf{v}_k\mathbf{v}_k^H
+
\sigma_r^2\Vert\mathbf{u}_k^H\Vert^2_2
]
\},\label{Phi_trans_coefficient}\\
&
c_{t,k} \triangleq \omega_k\vert\beta_k\vert^2\sigma_t,
c_{1,1}
\triangleq {\sum}_{k=1}^{K}  c_{t,k}\Vert\mathbf{a}_{1,k}\Vert^2_2,
c_{1} \triangleq c_{1,0} + c_{1,1},
\mathbf{t}_{1,1}
\triangleq
{\sum}_{k=1}^{K}-
c_{t,k}(b_{1,k}\mathbf{B}^{\ast}\mathbf{a}_{1,k}^{\ast}),
\mathbf{t}_{1,2}
\triangleq
{\sum}_{k=1}^{K}-
c_{t,k}\mathbf{b}_{2,k}\mathbf{a}_{1,k}^H\mathbf{b}_3,\nonumber\\
&\mathbf{t}_{1,0}
\triangleq
{\sum}_{k=1}^{K}
\big\{
\omega_k\beta_k^{\ast}\sqrt{q_k}\mathbf{u}_k^H\mathbf{P}_k
-
\omega_k\vert\beta_k\vert^2
\big[
{\sum}_{i=1}^{K}q_i
(
\mathbf{h}_{BU,i}^H\mathbf{u}_k\mathbf{u}_k^H\mathbf{P}_{i})
+
\mathbf{v}_k^{\ast}\mathbf{S}_k^{\ast}
\big]
\big\}^H,
\mathbf{t}_1 \triangleq \mathbf{t}_{1,0} + \mathbf{t}_{1,1} + \mathbf{t}_{1,2},
c_{r} \triangleq \sigma_t/\Gamma_r,
\nonumber\\
& \mathbf{T}_{1,0} \triangleq
{\sum}_{k=1}^{K}
\big\{
\omega_k\vert\beta_k\vert^2
\big[
{\sum}_{i=1}^{K}q_i(\mathbf{P}_{i}^H\mathbf{u}_k\mathbf{u}_k^H\mathbf{P}_{i})
\!+\!
\mathbf{S}_k^T\mathbf{S}_k^{\ast}\big]\big\},
 \mathbf{T}_{1,1} \!\triangleq\!\! {\sum}_{k=1}^{K}\! c_{t,k}(\vert b_{1,k}\vert^2\mathbf{B}^{*}\mathbf{B}^T),
 \mathbf{T}_{1,2} \!\triangleq\!\! {\sum}_{k=1}^{K} \! c_{t,k}((\mathbf{B}\mathbf{b}_3)^T\!\!\otimes\!({b}_{1,k}^{\ast}\mathbf{b}_{2,k})),\nonumber\\
& \mathbf{T}_{1,3} \triangleq {\sum}_{k=1}^{K} c_{t,k}((\mathbf{b}_3^H\mathbf{B}^H)^T\otimes({b}_{1,k}\mathbf{b}_{2,k}^H)),
 \mathbf{T}_{1,4} \triangleq {\sum}_{k=1}^{K} c_{t,k}((\mathbf{b}_3^H\mathbf{b}_3^H)^T\otimes(\mathbf{b}_{2,k}\mathbf{b}_{2,k}^H)),
 \mathbf{T}_{1} \triangleq \mathbf{T}_{1,0} + \mathbf{T}_{1,1} + \mathbf{T}_{1,2} + \mathbf{T}_{1,3} + \mathbf{T}_{1,4},\nonumber\\
&\mathbf{T}_{1,5} \triangleq {\sum}_{k=1}^{K} c_{t,k}\mathbf{b}_{2,k}\mathbf{a}_{1,k}^H\mathbf{B}^H,
\mathbf{T}_{1,6} \triangleq {\sum}_{k=1}^{K} c_{t,k}((\mathbf{B}\mathbf{B}^H)^T\otimes({b}_{1,k}^{\ast}\mathbf{b}_{2,k})),
\mathbf{T}_{1,7} \triangleq {\sum}_{k=1}^{K} c_{t,k}((\mathbf{b}_3^H\mathbf{B}^H)^T\otimes(\mathbf{b}_{2,k}\mathbf{b}_{2,k}^H)),\nonumber\\
&\mathbf{T}_{1,67} \triangleq \mathbf{T}_{1,6} + \mathbf{T}_{1,7},
\mathbf{T}_{1,8} \triangleq {\sum}_{k=1}^{K} c_{t,k}
( (\mathbf{B}\mathbf{B}^H)^T\otimes(\mathbf{b}_{2,k}\mathbf{b}_{2,k}^H)),
 c_{2,0}
\triangleq
{\sum}_{k=1}^{K}q_k\mathbf{u}_0^H\mathbf{h}_{BU,k}\mathbf{h}_{BU,k}^H\mathbf{u}_0
+
\mathbf{v}_0\mathbf{v}^H_0+\sigma_r^2\Vert\mathbf{u}_0^H\Vert^2_2,\nonumber\\
& c_{2,1}
\triangleq - c_{r}\Vert\mathbf{a}_{0}\Vert^2_2, c_{2} \triangleq c_{2,0} + c_{2,1},
 \mathbf{t}_{0,0} \triangleq -\big({\sum}_{k=1}^{K}q_k
(
\mathbf{h}_{BU,k}^H\mathbf{u}_0\mathbf{u}_0^H\mathbf{P}_{k})
+
\mathbf{v}_0^{\ast}\mathbf{S}_0^{\ast}\big)^H,
 \mathbf{t}_{0,1} \triangleq c_{r}(b_{0,1}\mathbf{B}^{\ast}\mathbf{a}_{0}^{\ast}),
 \mathbf{t}_{0,2} \triangleq c_{r}(\mathbf{b}_{0,2}\mathbf{a}_{0}^H\mathbf{b}_3),\nonumber \\
&\mathbf{t}_{0} = \mathbf{t}_{0,0} + \mathbf{t}_{0,1} + \mathbf{t}_{0,2},
\mathbf{T}_0
\triangleq
\big({\sum}_{k=1}^{K}q_k\mathbf{P}_{k}^H\mathbf{u}_0\mathbf{u}_0^H\mathbf{P}_{k}
+
\mathbf{S}_0^T\mathbf{S}_0^{\ast}\big),
\mathbf{T}_{0,1}
\triangleq
c_{r}
(\vert b_{0,1}\vert^2\mathbf{B}^{*}\mathbf{B}^T),
\mathbf{T}_{0,2}
\triangleq c_{r}
(\mathbf{B}\mathbf{b}_3)^T\otimes(\mathbf{b}_{0,2}{b}_{0,1}^H),\nonumber\\
&\mathbf{T}_{0,3}
\triangleq c_{r}
(\mathbf{b}_3^H\mathbf{B}^H)^T\otimes({b}_{0,1}\mathbf{b}_{0,2}^H),
\mathbf{T}_{0,4}
\triangleq c_{r}
(\mathbf{b}_3^H\mathbf{b}_3^H)^T\otimes(\mathbf{b}_{0,2}\mathbf{b}_{0,2}^H),
\mathbf{T}_{0,0} \triangleq \mathbf{T}_{0,1} + \mathbf{T}_{0,2} + \mathbf{T}_{0,3} + \mathbf{T}_{0,4},
\mathbf{T}_{0,5} \triangleq c_{r}\mathbf{b}_{0,2}\mathbf{a}_{0}^H\mathbf{B}^H,\nonumber\\
&\mathbf{T}_{0,6} \triangleq c_{r}((\mathbf{B}\mathbf{B}^H)^T\otimes({b}_{0,1}^{\ast}\mathbf{b}_{0,2})),
\mathbf{T}_{0,7} \triangleq c_{r}((\mathbf{b}_3^H\mathbf{B}^H)^T\otimes(\mathbf{b}_{0,2}\mathbf{b}_{0,2}^H)),
\mathbf{T}_{0,67} \triangleq \mathbf{T}_{0,6} + \mathbf{T}_{0,7},
\mathbf{T}_{0,8} \triangleq c_{r}
( (\mathbf{B}\mathbf{B}^H)^T\otimes(\mathbf{b}_{0,2}\mathbf{b}_{0,2}^H)).\nonumber
\end{align}
\end{small}
\boldsymbol{\hrule}
\end{figure*}

Based on the above transformation,
the reflection phase shift  optimization reduces to solving the following problem
\begin{subequations}
\begin{align}
\textrm{(P2)}:\mathop{\textrm{min}}
\limits_{
\bm{\phi}
}\
& \bm{\phi}^H\mathbf{T}\bm{\phi}
-2\textrm{Re}\{\mathbf{t}_1^H\bm{\phi}\}+c_1\label{P2_obj}\\
&+2\textrm{Re}\{\bm{\phi}^H\mathbf{T}_{1,5}\bm{\phi}^{\ast}
+\textrm{vec}(\bm{\phi}\bm{\phi}^T)^H\mathbf{T}_{1,67}\bm{\phi}\}\nonumber\\
&+\textrm{vec}(\bm{\phi}\bm{\phi}^T)^H\mathbf{T}_{1,8}\textrm{vec}(\bm{\phi}\bm{\phi}^T)\nonumber\\
\textrm{s.t.}\
&\bm{\phi}^H\mathbf{T}_0\bm{\phi}
-2\textrm{Re}\{\mathbf{t}_0^H\bm{\phi}\}+c_2
-\bm{\phi}^H\mathbf{T}_{0,0}\bm{\phi}\label{P2_c1}\\
&-2\mathrm{Re}\{\bm{\phi}^H\mathbf{T}_{0,5}\bm{\phi}^{\ast}
+\textrm{vec}(\bm{\phi}\bm{\phi}^T)^H\mathbf{T}_{0,67}\bm{\phi}\}\nonumber\\
&-\textrm{vec}(\bm{\phi}\bm{\phi}^T)^H\mathbf{T}_{0,8}\textrm{vec}(\bm{\phi}\bm{\phi}^T)\leq 0,\nonumber\\
&\vert\phi_{m}\vert=1, \forall m \in \mathcal{M}\label{Phi_c_module_c}.
\end{align}
\end{subequations}

As seen above, the problem (P2) is highly challenging due to the presence of the quartic and cubic terms
in the objective (\ref{P2_obj}) and the constraint (\ref{P2_c1}).
Note these high-order terms indeed stem from the propagation channel of the radar probing signals, i.e.,
$\mathbf{H}(\bm{\Phi})\triangleq
(\mathbf{g}_r+\mathbf{G}_r^H\bm{\Phi}\mathbf{g}_{RT})(\mathbf{g}_t^H+\mathbf{g}_{RT}^T\bm{\Phi}\mathbf{G}_t)$,
which gives birth to the aforementioned quartic and cubic terms in the SINR functions of the communication and radar sensing signals.
To resolve the above difficulty,
it is highly desirable that we could somewhat ``dissolve'' the high-order terms,
e.g., reducing the problem from quartic to quadratic,
which is much more tractable.
In fact,
this could be fulfilled via splitting the term $\bm{\Phi}$ in $\mathbf{H}(\bm{\Phi})$.
Specifically, via introducing a copy $\bm{\Psi}_1$ of $\bm{\Phi}$ into $\mathbf{H}(\bm{\Phi})$,
i.e., we could equivalently rewrite the $\mathbf{H}(\bm{\Phi})$ as follows
\begin{align}
 \mathbf{H}(\bm{\Phi},\bm{\Psi}_1)=
(\mathbf{g}_r+\mathbf{G}_r^H\bm{\Psi}_1\mathbf{g}_{RT})(\mathbf{g}_t^H+\mathbf{g}_{RT}^T\bm{\Phi}\mathbf{G}_t), \label{P123_H}
\end{align}
where $\bm{\Psi}_1$ is indeed a copy of $\bm{\Phi}$,
i.e., $\bm{\Phi}=\bm{\Psi}_1$.
In fact,
via introducing the intermediate variable $\bm{\Psi}_1$,
optimizing either $\bm{\Phi}$ or $\bm{\Psi}_1$ separately with the other one fixed will yield a quadratic problem.

Besides,
to decouple the non-convex constant magnitude constraints (\ref{Phi_c_module_c}),
we introduce another copy $\bm{\Psi}_2$ of $\bm{\Phi}$,
which will yield simple update (as will clear shortly).
Based on the discussions,
by introducing the two copies $\bm{\Psi}_1$ and $\bm{\Psi}_2$ of $\bm{\Phi}$ as above,
the problem (P2) can be equivalently written as
\begin{subequations}
\begin{align}
&{\textrm{(P3)}:  \mathop{\textrm{min}}
\limits_{
\bm{\phi},
\bm{\psi}_1,
\bm{\psi}_2
}\
\bm{\phi}^H(\mathbf{T}_{1,0}+\mathbf{T}_{1,1} )\bm{\phi}
+ \bm{\psi}_1^H\mathbf{T}_{1,2}\bm{\phi}}\\
&{+\! \bm{\phi}^H\mathbf{T}_{1,3}\bm{\psi}_1
\!+\! \bm{\psi}_1^H\mathbf{T}_{1,4}\bm{\psi}_1
\!-\!2\textrm{Re}\{(\mathbf{t}_{1,0}\! +\! \mathbf{t}_{1,1} )^H\bm{\phi}\!+\!\mathbf{t}_{1,2}^H\bm{\psi}_1\!\}}\nonumber\\
&{+ 2\textrm{Re}\{\bm{\psi}_1^H\mathbf{T}_{1,5}\bm{\phi}^{\ast}
+\textrm{vec}(\bm{\psi}_1\bm{\phi}^T)^H(\mathbf{T}_{1,6}\bm{\phi}+\mathbf{T}_{1,7}\bm{\psi}_1)\}}\nonumber\\
&{+ \textrm{vec}(\bm{\psi}_1\bm{\phi}^T)^H\mathbf{T}_{1,8}\textrm{vec}(\bm{\psi}_1\bm{\phi}^T)
+c_1}\nonumber\\
&\textrm{s.t.}\
 \bm{\phi}^H\mathbf{T}_{0,0}\bm{\phi}
-2\textrm{Re}\{(\mathbf{t}_{0,0} + \mathbf{t}_{0,1} )^H\bm{\phi}+\mathbf{t}_{0,2}^H\bm{\psi}_1\}\\
&-\bm{\phi}^H\mathbf{T}_{0,1} \bm{\phi}
- \bm{\psi}_1^H\mathbf{T}_{0,2}\bm{\phi}
- \bm{\phi}^H\mathbf{T}_{0,3}\bm{\psi}_1
- \bm{\psi}_1^H\mathbf{T}_{0,4}\bm{\psi}_1\nonumber\\
&- 2\textrm{Re}\{\bm{\psi}_1^H\mathbf{T}_{0,5}\bm{\phi}^{\ast}
+\textrm{vec}(\bm{\psi}_1\bm{\phi}^T)^H(\mathbf{T}_{0,6}\bm{\phi}+\mathbf{T}_{0,7}\bm{\psi}_1)\}\nonumber\\
&- \textrm{vec}(\bm{\psi}_1\bm{\phi}^T)^H\mathbf{T}_{0,8}\textrm{vec}(\bm{\psi}_1\bm{\phi}^T)
+c_2\leq0\nonumber\\
& \bm{\phi} = \bm{\psi}_1,\ \bm{\phi} = \bm{\psi}_2,\label{P3_equality}\\
&\vert\psi_{2,m}\vert=1, \forall m \in \mathcal{M}.
\end{align}
\end{subequations}

To solve the above problem (P3),
following the PDD framework \cite{ref33},
we turn to optimize its augmented Lagrangian (AL) problem, given as follows
\begin{subequations}
\begin{align}
&{ \textrm{(P4)}:  \mathop{\textrm{min}}
\limits_{
\bm{\phi},
\bm{\psi}_1,
\bm{\psi}_2,
\bm{\lambda}_1,
\bm{\lambda}_2,
}\
\bm{\phi}^H(\mathbf{T}_{1,0}+\mathbf{T}_{1,1} )\bm{\phi}
+ \bm{\psi}_1^H\mathbf{T}_{1,2}\bm{\phi}}\\
&{\!+\! \bm{\phi}^H\mathbf{T}_{1,3}\bm{\psi}_1
\!\!+\!\! \bm{\psi}_1^H\mathbf{T}_{1,4}\bm{\psi}_1
\!\!-\!\!2\textrm{Re}\{(\mathbf{t}_{1,0}\! +\! \mathbf{t}_{1,1} )^H\bm{\phi}\!+\!\mathbf{t}_{1,2}^H\bm{\psi}_1\!\}}\nonumber\\
&{+ 2\textrm{Re}\{\bm{\psi}_1^H\mathbf{T}_{1,5}\bm{\phi}^{\ast}
+\textrm{vec}(\bm{\psi}_1\bm{\phi}^T)^H(\mathbf{T}_{1,6}\bm{\phi}+\mathbf{T}_{1,7}\bm{\psi}_1)\}}\nonumber\\
&{+ \textrm{vec}(\bm{\psi}_1\bm{\phi}^T)^H\mathbf{T}_{1,8}\textrm{vec}(\bm{\psi}_1\bm{\phi}^T)
+c_1}\nonumber\\
&{+
\frac{1}{2\rho}\Vert\bm{\phi}-\bm{\psi}_1\Vert_2^2
+
\textrm{Re}\{\bm{\lambda}_1^H(\bm{\phi}-\bm{\psi}_1)\}}\nonumber\\
&{+
\frac{1}{2\rho}\Vert\bm{\phi}-\bm{\psi}_2\Vert_2^2
+
\textrm{Re}\{\bm{\lambda}_2^H(\bm{\phi}-\bm{\psi}_2)\}}\nonumber\\
&\textrm{s.t.}\
 \bm{\phi}^H\mathbf{T}_{0,0}\bm{\phi}
-2\textrm{Re}\{(\mathbf{t}_{0,0} + \mathbf{t}_{0,1} )^H\bm{\phi}+\mathbf{t}_{0,2}^H\bm{\psi}_1\}\\
&-\bm{\phi}^H\mathbf{T}_{0,1} \bm{\phi}
- \bm{\psi}_1^H\mathbf{T}_{0,2}\bm{\phi}
- \bm{\phi}^H\mathbf{T}_{0,3}\bm{\psi}_1
- \bm{\psi}_1^H\mathbf{T}_{0,4}\bm{\psi}_1\nonumber\\
&- 2\textrm{Re}\{\bm{\psi}_1^H\mathbf{T}_{0,5}\bm{\phi}^{\ast}
+\textrm{vec}(\bm{\psi}_1\bm{\phi}^T)^H(\mathbf{T}_{0,6}\bm{\phi}+\mathbf{T}_{0,7}\bm{\psi}_1)\}\nonumber\\
&- \textrm{vec}(\bm{\psi}_1\bm{\phi}^T)^H\mathbf{T}_{0,8}\textrm{vec}(\bm{\psi}_1\bm{\phi}^T)
+c_2\leq0\nonumber\\
&\vert\psi_{2,m}\vert=1, \forall m \in \mathcal{M}.
\end{align}
\end{subequations}

Guided by the PDD framework \cite{ref33},
we conduct a two-layer iteration procedure,
with its inner layer  updating $\bm{\phi}$, $\bm{\psi}_1$ and $\bm{\psi}_2$
in a block coordinate descent (BCD) manner
and its outer layer selectively updating the penalty coefficient $\rho$ or the dual variables $\{\bm{\lambda}_1, \bm{\lambda}_2\}$.
The PDD  procedure will be elaborated in the following.

\normalem
\underline{\emph{Inner Layer Procedure}}

For the inner layer iteration,
we will update $\bm{\phi}$, $\bm{\psi}_1$ and $\bm{\psi}_2$ in sequence.
When $\{\bm{\psi}_1, \bm{\psi}_2\}$ are given,
the minimization of AL with respect to (w.r.t.) $\bm{\phi}$ reduces to solving the following problem
\begin{subequations}
\begin{align}
\textrm{(P5)}:\mathop{\textrm{min}}
\limits_{
\bm{\phi}
}\
& \bm{\phi}^H\mathbf{\hat{T}}_1\bm{\phi}
-2\textrm{Re}\{\mathbf{\hat{t}}_1^H\bm{\phi}\}+\hat{c}_1\\
&+
\frac{1}{2\rho}\Vert\bm{\phi}-\bm{\psi}_1\Vert_2^2
+
\textrm{Re}\{\bm{\lambda}_1^H(\bm{\phi}-\bm{\psi}_1)\}\nonumber\\
&+
\frac{1}{2\rho}\Vert\bm{\phi}-\bm{\psi}_2\Vert_2^2
+
\textrm{Re}\{\bm{\lambda}_2^H(\bm{\phi}-\bm{\psi}_2)\}\nonumber
\\
\mathrm{s.t.}\
\bm{\phi}^H\mathbf{T}_{0,0}&\bm{\phi}
-2\textrm{Re}\{\mathbf{\hat{t}}_0^H\bm{\phi}\}+\hat{c}_2-\bm{\phi}^H\mathbf{{T}}_{0,9}\bm{\phi}\leq 0,\label{P5_c}
\end{align}
\end{subequations}
where the newly introduced coefficients are defined as follows
\begin{small}
\begin{align}
& \mathbf{h}_{\psi_1} \triangleq \mathbf{g}_r + \mathbf{G}_r^H\textrm{diag}(\bm{\psi}_1)\mathbf{g}_{RT},
 \mathbf{G}_{t1} \triangleq \textrm{diag}(\mathbf{g}_{RT}^T)\mathbf{G}_t,\\
& \mathbf{t}_{1,3} \triangleq  -{\sum}_{k=1}^{K}c_{t,k}
(\mathbf{G}_{t1}^{\ast}\mathbf{W}^{\ast}\mathbf{W}^T\mathbf{g}_t^{\ast}\mathbf{h}_{\bm{\psi}_1}^T
\mathbf{u}_k^{\ast}\mathbf{u}_{k}^T\mathbf{h}_{\bm{\psi}_1}^{\ast}),\nonumber\\
&\mathbf{T}_{1,9} \triangleq {\sum}_{k=1}^{K}c_{t,k}
( ( \mathbf{G}_{t1}\mathbf{W}\mathbf{W}^H\mathbf{G}_{t1}^H )^T\otimes( \mathbf{h}_{\bm{\psi}_1}\mathbf{u}_k\mathbf{u}_k^H\mathbf{h}_{\bm{\psi}_1}^H )  ),\nonumber\\
&{ c_{1,2}\! \triangleq\!\! {\sum}_{k=1}^{K}c_{t,k}\Vert\mathbf{u}_k^H\mathbf{h}_{\psi_1}\mathbf{g}_t^H\mathbf{W}\Vert_2^2,
c_{2,2}\! \triangleq\!
\!-\!c_{r}  \Vert\mathbf{u}_0^H\mathbf{h}_{\psi_1}\mathbf{g}_t^H\mathbf{W}\Vert_2^2,}\nonumber\\
& \mathbf{t}_{0,3}
\triangleq
c_{r}
(\mathbf{G}_{t1}^{\ast}\mathbf{W}^{\ast}\mathbf{W}^T\mathbf{g}_t^{\ast}\mathbf{h}_{\bm{\psi}_1}^T
\mathbf{u}_0^{\ast}\mathbf{u}_{0}^T\mathbf{h}_{\bm{\psi}_1}^{\ast}),\nonumber\\
&\mathbf{T}_{0,9} \triangleq
c_{r}  ( ( \mathbf{G}_{t1}\mathbf{W}\mathbf{W}^H\mathbf{G}_{t1}^H )^T\otimes( \mathbf{h}_{\bm{\psi}_1}\mathbf{u}_0\mathbf{u}_0^H\mathbf{h}_{\bm{\psi}_1}^H )  ),\nonumber\\
& \hat{c}_1 \triangleq c_{1,0} + c_{1,2},
 \mathbf{\hat{t}}_1 \triangleq \mathbf{t}_{1,0} + \mathbf{t}_{1,3},
 \mathbf{\hat{T}}_{1} \triangleq \mathbf{{T}}_{1,0} + \mathbf{{T}}_{1,9},\nonumber\\
& \hat{c}_2 \triangleq c_{2,0} + c_{2,2},
 \mathbf{\hat{t}}_0 \triangleq \mathbf{t}_{0,0} + \mathbf{t}_{0,3}.\nonumber
\end{align}
\end{small}

Obviously,
the problem (P5) is non-convex since the  constraint (\ref{P5_c}) is nonconvex.
We adopt the MM framework \cite{ref32} to convexify (\ref{P5_c}) via taking linearization of convex
terms at the point of $\bm{\phi}_0$,
which is given as
\begin{align}
\bm{\phi}^H\mathbf{{T}}_{0,9}\bm{\phi}
\geq
2\textrm{Re}\{ \bm{\phi}^H_0\mathbf{{T}}_{0,9}(\bm{\phi}-\bm{\phi}_0)\}
+
\bm{\phi}_0^H\mathbf{{T}}_{0,9}\bm{\phi}_0.\label{P5_c_SCA}
\end{align}

Therefore,
we turn to replace the term $\bm{\phi}^H\mathbf{{T}}_{0,9}\bm{\phi}$ in constraint (\ref{P5_c}) by (\ref{P5_c_SCA}),
and the problem (P5) is rewritten as
\begin{subequations}
\begin{align}
\textrm{(P6)}:\mathop{\textrm{min}}
\limits_{
\bm{\phi}
}\
& \bm{\phi}^H\mathbf{\hat{T}}_1\bm{\phi}
-2\textrm{Re}\{\mathbf{\hat{t}}_1^H\bm{\phi}\}+\hat{c}_1\\
&+
\frac{1}{2\rho}\Vert\bm{\phi}-\bm{\psi}_1\Vert_2^2
+
\textrm{Re}\{\bm{\lambda}_1^H(\bm{\phi}-\bm{\psi}_1)\}\nonumber\\
&+
\frac{1}{2\rho}\Vert\bm{\phi}-\bm{\psi}_2\Vert_2^2
+
\textrm{Re}\{\bm{\lambda}_2^H(\bm{\phi}-\bm{\psi}_2)\}\nonumber
\\
\mathrm{s.t.}\
&\bm{\phi}^H\mathbf{T}_{0,0}\bm{\phi}
-2\textrm{Re}\{\mathbf{\acute{t}}_0^H\bm{\phi}\}+\acute{c}_2\leq 0,
\end{align}
\end{subequations}
where
$\mathbf{\acute{t}}_0\triangleq\mathbf{\bar{t}}_0 + \mathbf{{T}}_{0,9}^H\bm{\phi}_0$
and
$\acute{c}_2 \triangleq \bar{c}_2 + (\bm{\phi}_0^H\mathbf{{T}}_{0,9}\bm{\phi}_0)^{\ast} $.
The problem (P6)
is a typical second order cone program (SOCP) and can
be solved by existing convex optimization solvers, e.g., CVX \cite{ref34}.

Given the  variables $\{\bm{\phi}, \bm{\psi}_2 \}$,
the optimization of updating  the auxiliary variable $\bm{\psi}_1$ is
formulated as
\begin{subequations}
\begin{align}
\textrm{(P7)}:\mathop{\textrm{min}}
\limits_{
\bm{\psi}_1
}\
& \bm{\psi}^H_1\mathbf{T}_{1,10}\bm{\psi}_1
-2\textrm{Re}\{\mathbf{t}^H_{1,3}\bm{\psi}_1\}+\tilde{c}_{1}\\
&+
\frac{1}{2\rho}\Vert\bm{\phi}-\bm{\psi}_1\Vert_2^2
+
\textrm{Re}\{\bm{\lambda}_1^H(\bm{\phi}-\bm{\psi}_1)\}\nonumber
\\
\mathrm{s.t.}\
&
-2\textrm{Re}\{\mathbf{t}_{0,4}^H\bm{\psi}_1\}+\tilde{c}_{2}-\bm{\psi}_1^H\mathbf{T}_{0,10}\bm{\psi}_1\leq 0,\label{P7_c}
\end{align}
\end{subequations}
where
the above newly introduced coefficients  defined as
\begin{small}
\begin{align}
&\mathbf{h}_2 \triangleq \mathbf{g}_t + \mathbf{G}_t^H\textrm{diag}(\bm{\phi})^H\mathbf{g}_{RT}^{\ast},
\mathbf{G}_{r1}\triangleq\textrm{diag}(\mathbf{g}_{RT}^H)\mathbf{G}_r,\\
&c_{1,3} \triangleq
{\sum}_{k=1}^{K}c_{k,t}\Vert\mathbf{u}_k^H\mathbf{g}_r\mathbf{h}_{2}^H\mathbf{W}\Vert_2^2,\nonumber\\
&  \mathbf{t}_{1,4}
\triangleq
-{\sum}_{k=1}^{K}c_{k,t}
(\mathbf{G}_{r1}\mathbf{u}_k\mathbf{u}_{k}^H\mathbf{g}_r\mathbf{h}_{2}^H\mathbf{W}\mathbf{W}^H\mathbf{h}_{2}),\nonumber\\
& \mathbf{T}_{1,10}
\triangleq
{\sum}_{k=1}^{K}c_{k,t}
(( \mathbf{h}_2^H\mathbf{W}\mathbf{W}^H\mathbf{h}_2 )^T\otimes( \mathbf{G}_{r1}\mathbf{u}_k\mathbf{u}_k^H\mathbf{G}_{r1}^H )),\nonumber\\
&c_{2,3} \triangleq
-c_{r} \Vert\mathbf{u}_0^H\mathbf{g}_r\mathbf{h}_{2}^H\mathbf{W}\Vert_2^2,\nonumber\\
&  \mathbf{t}_{0,4}
\triangleq
c_{r}
(\mathbf{G}_{r1}\mathbf{u}_0\mathbf{u}_{0}^H\mathbf{g}_r\mathbf{h}_{2}^H\mathbf{W}\mathbf{W}^H\mathbf{h}_{2}),\nonumber\\
& \mathbf{T}_{0,10}
\triangleq
c_{r}
(( \mathbf{h}_2^H\mathbf{W}\mathbf{W}^H\mathbf{h}_2 )^T\otimes( \mathbf{G}_{r1}\mathbf{u}_0\mathbf{u}_0^H\mathbf{G}_{r1}^H )),\nonumber\\
&\tilde{c}_1
\triangleq
\bm{\phi}^H\mathbf{T}_{1,0}\bm{\phi} - 2\textrm{Re}\{\mathbf{t}_{1,0}^H\bm{\phi}\}+c_{1,0}+c_{1,3},\nonumber\\
& \tilde{c}_2  \triangleq
 \bm{\phi}^H\mathbf{T}_{0,0}\bm{\phi} - 2\textrm{Re}\{\mathbf{t}_{0,0}^H\bm{\phi}\}+c_{2,0}+c_{2,3}\nonumber.
\end{align}
\end{small}

Obviously,
the non-convex constraint (\ref{P7_c})  makes the problem (P7) intractable.
Therefore,
still following the MM method,
we linearize the quadratic term $\bm{\psi}_1^H\mathbf{T}_{0,10}\bm{\psi}_1$ to obtain a tight lower bound as follows
\begin{align}
{
\!\!\!\!\!\bm{\psi}_1^H\!\mathbf{T}_{0,10}\bm{\psi}_1
\!\!\geq\!\!
2\textrm{Re}\{\! \bm{\psi}^H_{1,0}\!\mathbf{T}_{0,10}\!(\bm{\psi}_1\!\!-\!\bm{\psi}_{1,0}\!)\!\}
\!\!+\!\!
\bm{\psi}^H_{1,0}\!\mathbf{T}_{0,10}\bm{\psi}_{1,0},\label{P7_c_SCA}}
\end{align}
where  $\bm{\psi}_{1,0}$ is the value obtained in the last iteration.
Therefore,
by replace the term $\bm{\psi}_1^H\mathbf{T}_{0,10}\bm{\psi}_1$ by (\ref{P7_c_SCA}),
the  problem (P7) can be  rewritten as
\begin{subequations}
\begin{align}
\textrm{(P8)}:\mathop{\textrm{min}}
\limits_{
\bm{\psi}_1
}\
& \bm{\psi}^H_1\mathbf{T}_{1,10}\bm{\psi}_1
-2\textrm{Re}\{\mathbf{t}^H_{1,3}\bm{\psi}_1\}+\tilde{c}_{1}\\
&+
\frac{1}{2\rho}\Vert\bm{\phi}-\bm{\psi}_1\Vert_2^2
+
\textrm{Re}\{\bm{\lambda}_1^H(\bm{\phi}-\bm{\psi}_1)\}\nonumber
\\
\mathrm{s.t.}\
&
-2\textrm{Re}\{\mathbf{\tilde{t}}_{0,4}^H\bm{\psi}_1\}+\grave{c}_2\leq 0,
\end{align}
\end{subequations}
where
$ \mathbf{\tilde{t}}_{0,4}\! \triangleq\! \mathbf{{t}}_{0,4}\! +\! \mathbf{T}_{0,10}^H\bm{\psi}_{1,0} $
and
$ \grave{c}_2\! \triangleq\! \tilde{c}_{2} + (\bm{\psi}_{1,0}^H\mathbf{T}_{0,10}\bm{\psi}_{1,0})^{\ast} $.
The problem
(P8) is also an SOCP and solved by CVX.

When $\{\bm{\phi}, \bm{\psi}_1 \}$ are fixed,
the update of the auxiliary variable $\bm{\psi}_2$ is meant to solve
\begin{subequations}
\begin{align}
\textrm{(P9)}:\mathop{\textrm{min}}
\limits_{
\bm{\psi}
}\
&
\frac{1}{2\rho}\Vert\bm{\phi}-\bm{\psi}_2\Vert_2^2
+
\textrm{Re}\{\bm{\lambda}_2^H(\bm{\phi}-\bm{\psi})\}\label{Psi_obj}
\\
\textrm{s.t.}\
&\vert\psi_{2,m}\vert=1, \forall m \in \mathcal{M}.
\end{align}
\end{subequations}

Since $\bm{\psi}_2$ has unit modulus entries,
the quadratic term with respect to $\bm{\psi}_2$ in the objective function (\ref{Psi_obj})
is constant,
i.e., $\Vert\bm{\psi}_2\Vert^2_2/(2\rho)\!=\!M/(2\rho)$.
Hence the problem (P9) is reduced to
\begin{subequations}
\begin{align}
\textrm{(P10)}:\mathop{\textrm{max}}
\limits_{
\vert\bm{\psi}_2\vert=\mathbf{1}_M
}\
&\textrm{Re}\{(\bm{\phi}+\rho\bm{\lambda}_2)^H\bm{\psi}_2\}
\end{align}
\end{subequations}

Note that the maximum of problem (P10) can be readily achieved
when the elements of $\bm{\psi}_2$ are all aligned with those of the linear coefficient $(\rho^{-1}\bm{\phi}+\bm{\lambda}_2)$,
which is given as
\begin{align}
\bm{\psi}_2^{\star}=\textrm{exp}\big(j\cdot\angle(\bm{\phi}+\rho\bm{\lambda}_2)\big)\label{PSI_Solution}.
\end{align}

Since the inner layer update $\bm{\phi}$, $\bm{\psi}_1$ and  $\bm{\psi}_2$ in a BCD manner,
the objective value of (P4) will monotonically converge.

\underline{\emph{Outer Layer Procedure}}

When its convergence is  reached,
we adjust the value of dual variables $\{\bm{\lambda}_1, \bm{\lambda}_2\}$ or the penalty coefficient $\rho$ in the outer layer.
Specifically,
\begin{enumerate}[1)]
\item when the equations $\bm{\phi}=\bm{\psi}_1$ and $\bm{\phi}=\bm{\psi}_2$
are  approximately achieved,
i.e., $\Vert \bm{\phi}-\bm{\psi}_1\Vert _{\infty}$
and $\Vert \bm{\phi}-\bm{\psi}_2\Vert _{\infty}$ are simultaneously smaller than some predefined diminishing threshold
$\eta_k$ \cite{ref33},
then the dual variables $\bm{\lambda}_{\imath}$  will
 be  updated in a gradient ascent manner as follows:
\begin{align}
  \bm{\lambda}_{\imath}^{(k+1)}:=\bm{\lambda}_{\imath}^{(k)}+\rho^{-1}(\bm{\phi}-\bm{\psi}_{\imath}),\ {\imath} \in \{1, 2\};
\end{align}
\item when the equality constraints $\bm{\phi}=\bm{\psi}_1$
and/or
$\bm{\phi}=\bm{\psi}_2$  are far from ``being true",
in order to force $\bm{\phi}=\bm{\psi}_1$
and/or
$\bm{\phi}=\bm{\psi}_2$  being achieved in the subsequent iterations,
the outer layer will  increase the penalty parameter $\rho^{-1}$ as follows:
\begin{align}
	\big(\rho^{(k+1)}\big)^{-1}:=c^{-1}\cdot\big(\rho^{(k)}\big)^{-1},
\end{align}
where $c$ is a predetermined positive constant which is usually smaller than 1
and  typically chosen in the range of [0.8, 0.9] \cite{ref33}.
\end{enumerate}

The PDD-based method to solve problem (P2) is summarized in Algorithm \ref{alg:PDD}.
\begin{algorithm}[t]
\caption{PDD Method to Solve (P2)}
\label{alg:PDD}
\begin{algorithmic}[1]
\STATE {initialize}
$\bm{\phi}^{(0)}$,
$\bm{\psi}_{\imath}^{(0)}$,
$\bm{\lambda}_{\imath}^{(0)}$,
${\rho}^{(0)}$,
${\imath} \in \{1, 2\}$
and
$k=1$
;
\REPEAT
\STATE set $\bm{\phi}^{(k-1,0)}:=\bm{\phi}^{(k-1)}$,
$\bm{\psi}_{\imath}^{(k-1,0)}:=\bm{\psi}_{\imath}^{(k-1)}$,
$t=0$;
\REPEAT
\STATE  update $\bm{\phi}^{(k-1,t+1)}$  by  solving (P6);
\STATE  update $\bm{\psi}_{1}^{(k-1,t+1)}$  by  solving (P8);
\STATE  update $\bm{\psi}_2^{(k-1,t+1)}$ by (\ref{PSI_Solution});
\STATE  $t++$;
\UNTIL{$convergence$}
\STATE set $\bm{\phi}^{(k)}:=\bm{\phi}^{(k-1,\infty)}$,
$\bm{\psi}_{\imath}^{(k)}:=\bm{\psi}_{\imath}^{(k-1,\infty)}$;
\IF{$\Vert\bm{\phi}^{(k)}-\bm{\psi}_1^{(k)}\Vert_{\infty}\leq\eta_k $
and
$\Vert\bm{\phi}^{(k)}-\bm{\psi}_2^{(k)}\Vert_{\infty}\leq\eta_k $   }
\STATE{ $\bm{\lambda}_{\imath}^{(k+1)}:=\bm{\lambda}_{\imath}^{(k)}+\dfrac{1}{{\rho}^{(k)}}(\bm{\phi}^{(k)}-\bm{\psi}_{\imath}^{(k)})$, ${\rho}^{(k+1)}:= {\rho}^{(k)}$};
\ELSE
\STATE{$\bm{\lambda}_{\imath}^{(k+1)}:=\bm{\lambda}_{\imath}^{(k)}$, $1/{\rho}^{(k+1)}:= 1/(c\cdot{\rho}^{(k)})$};
\ENDIF \STATE $k++$;
\UNTIL{$\Vert\bm{\phi}^{(k)}-\bm{\psi}_1^{(k)}\Vert_{2}$
and
$\Vert\bm{\phi}^{(k)}-\bm{\psi}_2^{(k)}\Vert_{2}$
are sufficiently small simultaneously;}
\end{algorithmic}
\end{algorithm}

\subsection{Updating The BS Beamformer $\mathbf{W}$}
In this subsection, we discuss the update of the transmit beamformer $\mathbf{W}$.
With other variables being fixed, the optimization problem of updating $\mathbf{W}$ can be formulated as
\begin{subequations}
\begin{align}
\textrm{(P11)}:
\mathop{\textrm{min}}
\limits_{\mathbf{w}}\
&
\mathbf{w}^H\mathbf{D}_1\mathbf{w}-c_3\\
\textrm{s.t.}\
&\mathbf{w}^H\mathbf{D}_2\mathbf{w}-\mathbf{w}^H\mathbf{D}_3\mathbf{w}+c_4\leq0,\label{Beam_nonconvex_con}\\
&\mathbf{w}^H\mathbf{w}\leq P_{BS}.
\end{align}
\end{subequations}
with the new parameters defined as follows
\begin{align}
&\mathbf{w}\triangleq \textrm{vec}(\mathbf{W}),
c_4\triangleq
({\sum}_{k=1}^{K} q_k\vert\mathbf{u}^H_0\mathbf{h}_{U,k}\vert^2
\!+\!\sigma^2_r\Vert\mathbf{u}_0^H\Vert^2_2),\\
&{\mathbf{D}_2\!\triangleq \! \mathbf{I}_{N_t}\!\otimes\!\mathbf{G}^H\mathbf{u}_0\mathbf{u}_0^H\mathbf{G},
\mathbf{D}_3\!\triangleq\!  (\mathbf{I}_{N_t}\!\otimes\!\sigma_t^2\mathbf{H}^H\!\mathbf{u}_0\mathbf{u}_0^H\mathbf{H})/\Gamma_r,}\nonumber\\
&c_3=
\big[{\sum}_{k=1}^{K}\big(\mathrm{log}(\omega_k)
-\omega_k
+2\textrm{Re}\{\omega_k\beta_k^{\ast}\sqrt{q_k}\mathbf{u}^H_k\mathbf{h}_{U,k}\}\nonumber\\
&\quad-\omega_k\vert\beta_k\vert^2({\sum}_{i=1}^{K}q_i\vert\mathbf{u}^H_k\mathbf{h}_{U,i}\vert^2
+\sigma_r^2\Vert\mathbf{u}^H_k\Vert^2_2)+1\big)\big],\nonumber\\
&{\mathbf{D}_1\!\!\triangleq\!\!\! {\sum}_{k=1}^{K}\!\omega_k\vert\beta_k\vert^2
(\!\mathbf{I}_{N_t}\!\!\otimes\!\sigma_t^2\mathbf{H}^H\!\mathbf{u}_k\mathbf{u}_k^H\mathbf{H}
\!\!+\!\!
\mathbf{I}_{N_t}\!\!\otimes\!\mathbf{G}^H\!\mathbf{u}_k\mathbf{u}_k^H\!\mathbf{G}\!).}\nonumber
\end{align}

It can be observed that the problem (P11) is  still difficult to solve due to the difference of convex (DC) form constraint (\ref{Beam_nonconvex_con}).
Inspired by the MM framework,
we construct a linear lower-bound of the constraint (\ref{Beam_nonconvex_con}),
which is given as
\begin{align}
\mathbf{w}^H\mathbf{D}_3\mathbf{w}
\geq
2\textrm{Re}\{{\mathbf{w}}^H_0\mathbf{D}_3(\mathbf{w}-{\mathbf{w}_0})\}+{\mathbf{w}}^H_0\mathbf{D}_3{\mathbf{w}}_0,\label{Beam_nonconvex_con_MM}
\end{align}
where ${\mathbf{w}_0}$ is obtained from the last iteration.
Therefore, the nonconvex constraint (\ref{Beam_nonconvex_con}) can be replaced by (\ref{Beam_nonconvex_con_MM})
and the optimization problem (P11) is rewritten as
\begin{subequations}
\begin{align}
\textrm{(P12)}:\mathop{\textrm{min}}
\limits_{\mathbf{w}}\
&
\mathbf{w}^H\mathbf{D}_1\mathbf{w}-c_3\\
\textrm{s.t.}\
&\mathbf{w}^H\mathbf{D}_2\mathbf{w}-2\textrm{Re}\{\mathbf{d}_3^H\mathbf{w}\}+\hat{c}_4\leq0,\\
&\mathbf{w}^H\mathbf{w}\leq P_{BS},
\end{align}
\end{subequations}
where
$\mathbf{d}_3 \triangleq \mathbf{D}_3^H{\mathbf{w}}_0$
and
$\hat{c}_4\triangleq{c}_4+({\mathbf{w}}^H_0\mathbf{D}_3{\mathbf{w}}_0)^{\ast}$.
The problem (P12)  is  an SOCP and solved by CVX.

\subsection{Optimizing The User Transmission Power}
With other variables being given,
the optimization problem of all the users' transmission power $\{q_k\}$
can be formulated as
\begin{subequations}
\begin{align}
\textrm{(P13)}:\mathop{\textrm{min}}
\limits_{
\{q_k\}}\
&   {\sum}_{k=1}^{K}a_k q_k + {\sum}_{k=1}^{K}b_k \sqrt{q_k}-c_5\\
\textrm{s.t.}\
& {\sum}_{k=1}^{K}d_k q_k \leq \hat{c}_5,\\
& 0 \leq q_k\leq {P}_{U,k}, \forall k \in \mathcal{K},
\end{align}
\end{subequations}
where the newly introduced coefficients are defined as follows
\begin{align}
&a_k \triangleq {\sum}_{j=1}^{K}\omega_j\vert\beta_j\vert^2\vert\mathbf{u}_j^H\mathbf{h}_{U,k}\vert^2,
d_k \triangleq \vert\mathbf{u}^H_0\mathbf{h}_{U,k}\vert^2,\\
&\hat{c}_5 \triangleq  \sigma_t^2\Vert\mathbf{u}^H_0\mathbf{H}\mathbf{W}\Vert^2_2/\Gamma_r-\Vert\mathbf{u}^H_0\mathbf{G}\mathbf{W}\Vert^2_2
-\sigma^2_r\Vert\mathbf{u}_0^H\Vert^2_2,\nonumber\\
&b_k \triangleq -2\textrm{Re}(\omega_k\beta_k^{\ast}\mathbf{u}_k^H\mathbf{h}_{U,k}),
c_5
\triangleq
{\sum}_{k=1}^{K}
\big\{\mathrm{log}(\omega_k)
-\omega_k\nonumber\\
&-\omega_k\vert\beta_k\vert^2(\sigma_t^2\Vert\mathbf{u}_k^H\mathbf{H}\mathbf{W}\Vert^2_2
+\Vert\mathbf{u}_k^H\mathbf{G}\mathbf{W}\Vert^2_2
+\sigma_r^2\Vert\mathbf{u}_k^H\Vert)+1
\big\}.\nonumber
\end{align}

The problem (P13) can still be formulated into an SOCP problem and can be numerically solved.

\subsection{Optimizing The  Receiver Filter $\{\mathbf{u}_k\}$}
The update of $\{\mathbf{u}_k\}$ are meant to solve the following problem
\begin{subequations}
\begin{align}
\textrm{(P14)}:\mathop{\textrm{min}}
\limits_{\{\mathbf{u}_k\}}\
& {\sum}_{k=1}^{K}(\mathbf{u}_k^H\mathbf{F}_{k}\mathbf{u}_k
\!-\!
2\textrm{Re}\{\mathbf{u}^H_k\mathbf{\tilde{h}}_{U,k}\})
\!-\!c_6
\end{align}
\end{subequations}
where the above newly introduced coefficients are defined as
\begin{small}
\begin{align}
&\mathbf{\tilde{h}}_{U,k}\! \triangleq \!\omega_k\beta_k^{\ast}\sqrt{q_k}\mathbf{h}_{U,k},
c_6
\!\triangleq\!
{\sum}_{k=1}^{K}\big(
\textrm{log}(\omega_k)+\omega_k+1\big),\\
&{\mathbf{F}_{\!k}
\!\!\triangleq\!\!
\omega_{\!k}\!\vert\beta_k\vert^2\!
\big(\!{\sum}_{i=1}^{K}\!q_i\mathbf{h}_{U,i}\mathbf{h}_{U,i}^H
\!\!+\!\!\sigma_t^2\!\mathbf{H}\mathbf{W}\mathbf{W}^{\!H}\!\mathbf{H}^{\!H}\!
\!\!\!+\!\!\mathbf{G}\mathbf{W}\mathbf{W}^{\!H}\!\!\mathbf{G}^{\!H}\!
\!\!\!+\!\!\sigma_r^2\mathbf{I}_{\!N_r}\!\!\big).}\nonumber
\end{align}
\end{small}

It is obviously that the problem (P14) can be decomposed into $K$ independent sub-problems,
which each subproblem being given as
\begin{subequations}
\begin{align}
\textrm{(P}\mathrm{\textrm{15}_k}\textrm{)}:\mathop{\textrm{min}}
\limits_{\mathbf{u}_k}\
& \mathbf{u}_k^H\mathbf{F}_{k}\mathbf{u}_k
-
2\textrm{Re}\{\mathbf{u}^H_k\mathbf{\tilde{h}}_{U,k}\}
\end{align}
\end{subequations}

Notice that the problem  (P$\mathrm{\textrm{15}_k}$) is a typical unconstrained convex quadratic problem.
Its optimal solution can be easily obtained via setting its derivative to zero and obtained as follows
\begin{align}
\mathbf{u}_k^{\star}=\mathbf{F}_{k}^{-1}\mathbf{\tilde{h}}_{U,k}, \forall k \in \mathcal{K}.\label{Solution_u_k}
\end{align}

\subsection{Optimizing The Target Siganl Receiver Filter $\mathbf{u}_0$}
After fixing other variables,
the optimization problem w.r.t.
$\mathbf{u}_0$ is reduced to  a feasibility check problem
\begin{subequations}
\begin{align}
\textrm{(P16)}:{\textrm{Find}}\
& \mathbf{u}_0\\
\textrm{s.t.}\ &
\mathbf{u}_0^H\mathbf{E}_1\mathbf{u}_0-\mathbf{u}_0^H\mathbf{E}_2\mathbf{u}_0
\leq
0.\label{(EE_SINR)}
\end{align}
\end{subequations}
where the new parameters in the above are defined as
\begin{align}
&\mathbf{E}_2
\triangleq
\sigma_t^2\mathbf{H}\mathbf{W}\mathbf{W}^H\mathbf{H}^H/\Gamma_r,\\
&\mathbf{E}_1
\triangleq
\big({\sum}_{k=1}^{K}q_k\mathbf{h}_{U,k}\mathbf{h}_{U,k}^H+\mathbf{G}\mathbf{W}\mathbf{W}^H\mathbf{G}^H+\sigma_r^2\mathbf{I}_{N_r}\big).\nonumber
\end{align}

The feasibility characterization problem (P16),
whose objective is missing, is also known as Phase-I problem \cite{ref38}.
To solve it, we consider another closely related problem as follows
\begin{subequations}
\begin{align}
\textrm{(P17)}:\mathop{\textrm{min}}
\limits_{\mathbf{u}_0, \alpha_u}\
& \alpha_u\\
\textrm{s.t.}\
&\mathbf{u}_0^H\mathbf{E}_1\mathbf{u}_0-\mathbf{u}_0^H\mathbf{E}_2\mathbf{u}_0
\leq
\alpha_u\label{Radar_SINR_c}.
\end{align}
\end{subequations}

To see the connection between
(P16) and (P17),
suppose that the optimal solution to (P17) is
($\mathbf{u}_0^{\star}$,$\alpha_u^\star$).
If $\alpha_u^\star\leq0$,
then $\alpha_u^\star$
is actually a feasible solution to (P16).
Note that (P17) is assured to have a feasible solution yielding non-positive $\alpha_u$
if the whole iteration starts from a feasible point.
Minimizing (P17) is meant to find a more ``feasible'' $\mathbf{u}_0$,
which provides a larger margin to satisfy the constraint (\ref{(EE_SINR)}) and hence benefits the optimization of other variables.

Obviously,
the problem (P17) obtains optimality only
when (\ref{Radar_SINR_c}) achieves equality.
Therefore, solving (P17) is equivalent to minimizing the left hand side of (\ref{Radar_SINR_c}),
i.e., (\ref{P15_obj}).
\begin{align}
\textrm{(P18)}:\mathop{\textrm{min}}
\limits_{\mathbf{u}_0}\
& \mathbf{u}_0^H\mathbf{E}_1\mathbf{u}_0-\mathbf{u}_0^H\mathbf{E}_2\mathbf{u}_0\label{P15_obj}
\end{align}

Since the objective function (\ref{P15_obj}) is DC form,
we again adopt the MM method to convexify the term $-\mathbf{u}_0^H\mathbf{E}_2\mathbf{u}_0$ by  linearization as follows
\begin{align}
-\mathbf{u}_0^H\mathbf{E}_2\mathbf{u}_0
&\leq-
\hat{\mathbf{u}}_0^H\mathbf{E}_2\hat{\mathbf{u}}_0
-2\textrm{Re}\{\hat{\mathbf{u}}_0^H\mathbf{E}_2(\mathbf{u}_0-\hat{\mathbf{u}}_0)\}\label{U_0_linear}\\
&=
-2\textrm{Re}\{\hat{\mathbf{u}}_0^H\mathbf{E}_2\mathbf{u}_0\}
+
(\hat{\mathbf{u}}_0^H\mathbf{E}_2\hat{\mathbf{u}}_0)^{\ast},\nonumber
\end{align}
where $\hat{\mathbf{u}}_0$ is feasible solution obtained in the last iteration.

Therefore,
replacing the term $-\mathbf{u}_0^H\mathbf{E}_2\mathbf{u}_0$  by  (\ref{U_0_linear}),
we turn to optimize a tight convex upper bound of the objective (P18),
which is given as
\begin{align}
\textrm{(P19)}:\mathop{\textrm{min}}
\limits_{\mathbf{u}_0}\
& \mathbf{u}_0^H\mathbf{E}_1\mathbf{u}_0\!-\!2\textrm{Re}\{\hat{\mathbf{u}}_0^H\mathbf{E}_2\mathbf{u}_0\}
\!+\!
(\hat{\mathbf{u}}_0^H\mathbf{E}_2\hat{\mathbf{u}}_0)^{\ast}
\end{align}

The problem (P19) is also a unconstrained convex quadratic problem
and its optimal solution can be directly obtained as
\begin{align}
\mathbf{u}_0^{\star}
=
\mathbf{E}_1^{-1}(\mathbf{E}_2^H\hat{\mathbf{u}}_0).\label{Solution_u_0}
\end{align}

The overall algorithm to solve problem (P1) is  specified in Algorithm \ref{alg:Overall}.
\begin{algorithm}[t]
\caption{Proposed Algorithm to Solve (P1)}
\label{alg:Overall}
\begin{algorithmic}[1]
\STATE {randomly generate feasible}
$\bm{\phi}^{(0)}$,
$\mathbf{W}^{(0)}$,
$\{q_k^0\}$,
$\{\mathbf{u}_k\}$,
$\mathbf{u}_0$,
and
$i=0$;
\REPEAT
\STATE update $\{\beta_k^{(i+1)}\}$ and $\{\omega_k^{(i+1)}\}$ by (\ref{Solution_beta}) and (\ref{Solution_omega}), respectively.
\STATE update $\bm{\phi}^{(i+1)}$ by invoking Alg.1;
\STATE set $\mathbf{W}^{(i,0)} := \mathbf{W}^{(i)}$, $m=0$;
\REPEAT
\STATE update $\mathbf{W}^{(i,m+1)}$ by solving  (P12);
\STATE $m++$;
\UNTIL{$convergence$;}
\STATE set $\mathbf{W}^{(i+1)} := \mathbf{W}^{(i,\infty)}$;
\STATE update $\{{q}_k^{(i+1)}\}$ by solving (P13);
\STATE update $\{\mathbf{u}_k^{(i+1)}\}$ by (\ref{Solution_u_k});
\STATE set $\mathbf{u}_0^{(i,0)} := \mathbf{u}_0^{(i)}$, $m=0$;
\REPEAT
\STATE update $\mathbf{u}_0^{(i,m+1)}$ by (\ref{Solution_u_0});
\STATE $m++$;
\UNTIL{$convergence$;}
\STATE set $\mathbf{u}_0^{(i+1)} := \mathbf{u}_0^{(i,\infty)}$;
\STATE $i++$;
\UNTIL{$convergence$;}
\end{algorithmic}
\end{algorithm}

\section{Low-complexity Algorithm}

Note that our previously proposed Alg.2 relies on numerical solvers,
e.g., CVX,
to update various block coordinates,
including $\mathbf{W}$, $\bm{\phi}$ and $\{q_k\}$.
This feature may give rise to some undesirable properties:

i) general convex optimization solvers,
including CVX, relies on interior point (IP) method
\cite{ref352_convex}
to resolve SOCP problems,
whose complexity increases dramatically when variable's dimension grows.

ii) utilization of third-party solvers inevitably increases cost and inconvenience in implementing the algorithm,
e.g.,
purchase of license,
software installation/maintaince and the platform required to support the solver.

Therefore,
we proceed to explore solution that, hopefully, does not rely on any numerical solvers.

\subsection{Efficient Update of $\bm{\phi}$}
Firstly,
in order to  efficiently solve (P6) and (P8),
we introduce the following lemma that is proved in Appendix A.
\begin{lemma}
\label{lem:complex_Convex_trust_region problem}
Consider the following  problem:
\begin{subequations}
\begin{align}
(\textrm{P}_{Lm1}):\mathop{\textrm{min}}
\limits_{\mathbf{x}}\
& \mathbf{x}^H\mathbf{Q}\mathbf{x}-2\textrm{Re}\{\mathbf{q}^H\mathbf{x}\}+q\\
\textrm{s.t.}\
&\mathbf{x}^H\mathbf{\bar{Q}}\mathbf{x}-2\textrm{Re}\{\mathbf{\bar{q}}^H\mathbf{x}\}+\bar{q}\leq 0,
\end{align}
\end{subequations}
where $\mathbf{Q} \succ {0}$ and $\mathbf{\bar{Q}}\succcurlyeq {0}$,
and Slater's condition holds.
Then the optimal solution to problem ($\textrm{P}_{Lm1}$)
is given by one of the following two cases:
\begin{itemize}
\item[] \underline{CASE-I}:
If
$(\mathbf{Q}^{-1}\mathbf{q})^H\mathbf{\bar{Q}}(\mathbf{Q}^{-1}\mathbf{q})
-2\textrm{Re}\{\mathbf{\bar{q}}^H(\mathbf{Q}^{-1}\mathbf{q})\}+\bar{q}\leq 0$,
then the optimal solution $\mathbf{x}^{\star} = \mathbf{Q}^{-1}\mathbf{q}$.
\item[] \underline{CASE-II}:
Otherwise,
$\mathbf{x}^{\star} = (\varsigma^{\star}\mathbf{\bar{Q}}+\mathbf{Q})^{-1}(\varsigma^{\star}\mathbf{\bar{q}}+\mathbf{q})$,
where the positive $\varsigma^{\star}$ is the solution to the following equation
\begin{align}
&\big((\varsigma^{\star}\mathbf{\bar{Q}}+\mathbf{Q})^{-1}(\varsigma^{\star}\mathbf{\bar{q}}+\mathbf{q})\big)^H
\mathbf{\bar{Q}}
(\varsigma^{\star}\mathbf{\bar{Q}}+\mathbf{Q})^{-1}(\varsigma^{\star}\mathbf{\bar{q}}+\mathbf{q})\nonumber\\
&-2\textrm{Re}\{\mathbf{\bar{q}}^H(\varsigma^{\star}\mathbf{\bar{Q}}+\mathbf{Q})^{-1}(\varsigma^{\star}\mathbf{\bar{q}}+\mathbf{q})\}+\bar{q}= 0.
\end{align}
Note that the value of $\varsigma^{\star}$ can be efficiently obtained by the Newton's method.
\end{itemize}
\end{lemma}

Based on the above Lemma \ref{lem:complex_Convex_trust_region problem},
we examine the solution of (P6),
which is a preliminary step in PDD procedure.
For ease of notations,
we denote
\begin{align}
& \mathbf{T}_{1,11} \triangleq \mathbf{\hat{T}}_1 + \mathbf{I}/\rho,
 \mathbf{t}_{1,4} \triangleq \mathbf{\hat{t}}_1 + \frac{\bm{\psi}_1+\bm{\psi}_2}{2\rho} - \frac{\bm{\lambda}_1+\bm{\lambda}_2}{2},\\
& {c}_{1,4} \triangleq {\hat{c}}_1 + \frac{\Vert\bm{\psi}_1\Vert_2^2+\Vert\bm{\psi}_2\Vert_2^2}{2\rho} - \textrm{Re}\{\bm{\lambda}_1^H\bm{\psi}_1 + \bm{\lambda}_2^H\bm{\psi}_2 \}.\nonumber
\end{align}
Then we can rewrite the problem (P6)  as
\begin{subequations}
\begin{align}
\textrm{(P20)}:\mathop{\textrm{min}}
\limits_{
\bm{\phi}
}\
& \bm{\phi}^H\mathbf{T}_{1,11}\bm{\phi}
-2\mathrm{Re}\{\mathbf{t}_{1,4}^H\bm{\phi}\}+{c}_{1,4}\\
\mathrm{s.t.}\
&\bm{\phi}^H\mathbf{T}_{0,0}\bm{\phi}
-2\textrm{Re}\{\mathbf{\acute{t}}_0^H\bm{\phi}\}+\acute{c}_2\leq 0.
\end{align}
\end{subequations}

Obviously,
(P20) can be efficiently solved by Lemma \ref{lem:complex_Convex_trust_region problem}.

Next,
we turn to obtain  the solution of (P8) via Lemma \ref{lem:complex_Convex_trust_region problem}.
For simplicity,
we introduce the new definitions
\begin{align}
& \mathbf{T}_{1,12} \triangleq \mathbf{{T}}_{1,10} + \mathbf{I}/2\rho,
 \mathbf{t}_{1,5} \triangleq \mathbf{{t}}_{1,3} + \frac{\bm{\phi}}{2\rho} + \frac{\bm{\lambda}_1}{2},\\
& {c}_{1,5} \triangleq \tilde{c}_{1}
+ \frac{\Vert\bm{\phi}\Vert_2^2}{2\rho} + \textrm{Re}\{\bm{\lambda}_1^H\bm{\phi} \}.\nonumber
\end{align}
The problem (P8) can be equivalently rewritten as
 \begin{subequations}
\begin{align}
\textrm{(P21)}:\mathop{\textrm{min}}
\limits_{
\bm{\psi}_1
}\
& \bm{\psi}^H_1\mathbf{T}_{1,12}\bm{\psi}_1
-2\mathrm{Re}\{\mathbf{t}^H_{1,5}\bm{\psi}_1\}+{c}_{1,5}\\
\mathrm{s.t.}\
&
-2\textrm{Re}\{\mathbf{\tilde{t}}_{0,4}^H\bm{\psi}_1\}+\grave{c}_2\leq 0.
\end{align}
\end{subequations}
It is obvious that the above problem (P21) also can be easily solved by Lemma \ref{lem:complex_Convex_trust_region problem}.

\subsection{Efficient Update of $\mathbf{W}$}
In this subsection,
we investigate low complexity solution to the problem (P12) in Sec. III-B.
Firstly,
to develop analytic solution,
we introduce an auxiliary variable $\mathbf{f}$ and transform (P12) into an equivalent form as follows
\begin{subequations}
\begin{align}
\textrm{(P22)}:\mathop{\textrm{min}}
\limits_{\mathbf{w}, \mathbf{f}}\
&
\mathbf{w}^H\mathbf{D}_1\mathbf{w}-c_3\label{AEMM_oriPro_obj}\\
\textrm{s.t.}\
&\mathbf{f}^H\mathbf{D}_2\mathbf{f}-2\textrm{Re}\{\mathbf{d}_3^H\mathbf{f}\}+\hat{c}_4\leq0,\\
&\mathbf{w}^H\mathbf{w}\leq \textrm{P}_{BS},\\
&\mathbf{w} = \mathbf{f}\label{ADMM_equality_constraint}.
\end{align}
\end{subequations}

We then proceed via adopting the ADMM methodology \cite{ref35}
to solve the above problem.
Specifically,
by penalizing the equality constraint (\ref{ADMM_equality_constraint}) in the objective,
the AL problem of (P22) is given as
\begin{subequations}
\begin{align}
\textrm{(P23)}:\mathop{\textrm{min}}
\limits_{\mathbf{w}, \mathbf{f}, \bm{\tau}}\
&
\mathbf{w}^H\mathbf{D}_1\mathbf{w}\!+\!\textrm{Re}\{\bm{\tau}^H(\mathbf{w}\! -\! \mathbf{f})\}\!+\!\frac{\upsilon}{2}\Vert\mathbf{w}\! -\! \mathbf{f}\Vert^2_2\\
\textrm{s.t.}\
&\mathbf{f}^H\mathbf{D}_2\mathbf{f}-2\textrm{Re}\{\mathbf{d}_3^H\mathbf{f}\}+\hat{c}_4\leq0,\\
&\mathbf{w}^H\mathbf{w}\leq \textrm{P}_{BS},
\end{align}
\end{subequations}
where $\upsilon$ is a positive constant
and
$\bm{\tau}\in \mathbb{C}^{N_t\cdot N_t\times 1}$
is the introduced Lagrangian multiplier associated with (\ref{ADMM_equality_constraint}).
According to the ADMM method, to solve (P23),
we alternatively update $\mathbf{w}$, $\mathbf{f}$ and $\bm{\tau}$, as specified in the sequel.

With $\mathbf{w}$ and $\bm{\tau}$ being fixed,
the optimization of (P23) w.r.t. $\mathbf{f}$ is given as the following problem
\begin{subequations}
\begin{align}
\textrm{(P24)}:\mathop{\textrm{min}}
\limits_{\mathbf{f}}\
&
\Vert\mathbf{f}\Vert_2^2 - 2\textrm{Re}\{\mathbf{d}_1^H\mathbf{f}\}                      \\
\textrm{s.t.}\
&\mathbf{f}^H\mathbf{D}_2\mathbf{f}-2\textrm{Re}\{\mathbf{d}_3^H\mathbf{f}\}+\hat{c}_4\leq0,
\end{align}
\end{subequations}
where $\mathbf{d}_1 \triangleq \bm{\tau}/\upsilon+\mathbf{w}$.
It is readily seen that (P24) also satisfies the assumptions of Lemma \ref{lem:complex_Convex_trust_region problem}
and hence can be easily solved by the results therein.
Details are omitted to avoid repetition.

When $\mathbf{f}$ and $\bm{\tau}$ are given,
the update of $\mathbf{w}$ is reduced to solving the following problem
\begin{subequations}
\begin{align}
\textrm{(P25)}:\mathop{\textrm{min}}
\limits_{\mathbf{w}}\
&
\mathbf{w}^H\mathbf{\bar{D}}_1\mathbf{w}
-2\textrm{Re}\{\mathbf{\bar{d}}_1^H\mathbf{w}\}\\
\textrm{s.t.}\
&\mathbf{w}^H\mathbf{w}\leq P_{BS}.
\end{align}
\end{subequations}
where $\mathbf{\bar{D}}_1\triangleq \mathbf{D}_1+\frac{\upsilon}{2}\mathbf{I}_{N_t\cdot N_t} $
and  $ \mathbf{\bar{d}}_1\triangleq \frac{1}{2}({\upsilon}{\mathbf{f}}-\bm{\tau}) $.
Obviously,
the  problem (P25)
can also be efficiently solved
by exploiting  Lemma \ref{lem:complex_Convex_trust_region problem}.

Specifically,
the two CASEs identified in Lemma \ref{lem:complex_Convex_trust_region problem} can be accommodated to (P25) as follows
\begin{itemize}
\item[] {CASE-I}:
When
$(\mathbf{\bar{D}}^{-1}\mathbf{\bar{d}}_1)^H(\mathbf{\bar{D}}^{-1}\mathbf{\bar{d}}_1)
\leq \textrm{P}_0$,
the optimal solution $\mathbf{w}^{\star} = \mathbf{\bar{D}}^{-1}\mathbf{\bar{d}}_1$.
\item[] {CASE-II}:
Otherwise,
$\mathbf{w}^{\star} = (\mathbf{\bar{D}}+\kappa^{\star}\bm{I}_{N_t\cdot N_t})^{-1}\mathbf{\bar{d}}_1$.
The optimal value of  $\kappa^{\star}$  can be efficiently obtained
(e.g.,  Newton's method or bisection search).
\end{itemize}

Following the ADMM method \cite{ref35},
after the primal variables being updated,
the dual variable $\bm{\tau}$ is updated in a gradient ascent manner,
which is given as
\begin{align}
\bm{\tau}^{(t+1)} := \bm{\tau}^{(t)}+\upsilon(\mathbf{w}-\mathbf{f}).\label{ADMM_update_Lagrangian_multiplier}
\end{align}

The ADMM-based low complexity method to solve problem (P22), i.e., (P12), is summarized in Algorithm \ref{alg:ADMM}.
\begin{algorithm}[t]
\caption{ADMM Method to Solve (P22), i.e., (P12)}
\label{alg:ADMM}
\begin{algorithmic}[1]
\STATE {initialize}
$\mathbf{f}^{(0)}$,
$\mathbf{w}^{(0)}$,
$\bm{\tau}^{(0)}$
and
$t=0$
;
\REPEAT
\STATE update $\mathbf{f}^{(t+1)}$ by solving  (P24);
\STATE update $\mathbf{w}^{(t+1)}$ by solving  (P25);
\STATE update $\bm{\tau}^{(t+1)}$ by (\ref{ADMM_update_Lagrangian_multiplier});
\STATE $t++$;
\UNTIL{$convergence$;}
\end{algorithmic}
\end{algorithm}

\subsection{Efficient Update of $\{ q_k\}$}
In this subsection,
we develop closed-form solution to update $\{q_k\}$.
Firstly,
we define $p_k = \sqrt{q_k}$ and $\bar{P}_{U,k} = \sqrt{P_{U,k}}$,
and then the problem can be rewritten as
\begin{subequations}
\begin{align}
\textrm{(P26)}:\mathop{\textrm{min}}
\limits_{
\{p_k\}}\
&   {\sum}_{k=1}^{K}a_k p_k^2 + {\sum}_{k=1}^{K}b_k p_k-c_5\\
\textrm{s.t.}\
& {\sum}_{k=1}^{K}d_k p_k^2 \leq \hat{c}_5,\label{Q_sun_power_constraint}\\
& 0 \leq p_k\leq \bar{P}_{U,k}, \forall k \in \mathcal{K}.\label{Q_single_power_constraint}
\end{align}
\end{subequations}

The above problem
indeed has analytic solution as presented in the following theorem,
which is proved in Appendix B.

\begin{theorem}\label{alg:p_KKT}
We define $\hat{p}_k=\textrm{min}\{-\frac{b_k}{2a_k} ,\bar{P}_{U,k}\}$, $\forall k \in \mathcal{K}$.
The optimal values of $\{p_k\}$ are obtained according to one of the following
two cases:
\begin{itemize}
\item[] \underline{CASE-I}:
if ${\sum}_{k=1}^{K}d_k \hat{p}_k^2 \leq \hat{c}_5$, the optimal solutions $p_k^{\star} = \hat{p}_k$,
 $\forall k  \in \mathcal{K}$.
\item[] \underline{CASE-II}:
Otherwise, the optimal values of $\{p_k\}$ are given as
\begin{align}
p_k^{\star}(\nu^{\star}) = \bigg[-\frac{b_k}{2a_k + 2 \nu^{\star} d_k }\bigg]^{\bar{P}_{U,k}}, \forall k \in \mathcal{K},
\end{align}
where $[a]^{\bar{P}}\triangleq\textrm{min}\{a,\bar{P}\}$,
and $\nu^{\star}$  is the unique solution to the following equation
\begin{align}
{\sum}_{k=1}^{K}d_k \big({p}_k^{\star}(\nu^{\star})\big)^2 = \hat{c}_5.
\end{align}
\end{itemize}
\end{theorem}

The optimal value of $\nu^{\star}$ in \underline{CASE-II} of
Theorem \ref{alg:p_KKT} can be efficiently obtained by a bisection search procedure.
However,
the upper-bound of $\nu^{\star}$ is still missing.
Therefore,
the following lemma provides an upper-bound of $\nu^{\star}$,
which is proved in Appendix C
\begin{lemma}
\label{lem:The upper bound of nu}
Firstly, we assume that
$\{\tilde{p}_k\}$ is any strictly feasible solution of problem (P26)
and $\mathsf{obj}(\{\tilde{p}_k\})$ is the objective value of problem (P26) yielded by $\{\tilde{p}_k\}$.
Then,
following the definition of $\{\hat{p}_k\}$ introduced in Theorem \ref{alg:p_KKT},
an upper-bound of $\nu^{\star}$ is given as
\begin{align}
\nu^{\star}\leq \frac{\mathsf{obj}(\{\tilde{p}_k\})-\mathsf{obj}(\{\hat{p}_k\})}{\hat{c}_4-{\sum}_{k=1}^{K}d_k \tilde{p}_k^2}.\label{Lemma3_upper_bound}
\end{align}
\end{lemma}

\subsection{Complexity}
In the following,
we will discuss the complexity of our proposed algorithms.
According to the complexity analysis in \cite{ref36.1},
for the PDD-based algorithm,
the complexity of solving (P2) is $C_4C_3(C_1+C_2)M^3$,
where $C_1$ and $C_2$ denote the iteration number of solving (P6) and (P8) by MM method, respectively.
$C_3$ and $C_4$  represent the iteration number of  the outer and inner PDD loops, respectively.
In the each iteration,
the complexity of solving SOCP problems (P12) and (P13) are
$\mathcal{O}( C_5 N_t^6  )$
and
$\mathcal{O}( K^3 )$,
respectively,
where $C_5$ is the number of iterations of solving problem (P12) by the MM method.
Therefore,
the total computational complexity of Algorithm 2 is approximately given as
$\mathcal{O}(  C_6 (   C_4C_3(C_1\!+\!C_2)  M^3 + C_5 N_t^6 + K^3  )    )$
with $C_6$ represented as the number of iterations to solve problem (P2).

\section{Numerical Results}
\begin{figure}[t]
	\centering
	\includegraphics[width=.4\textwidth]{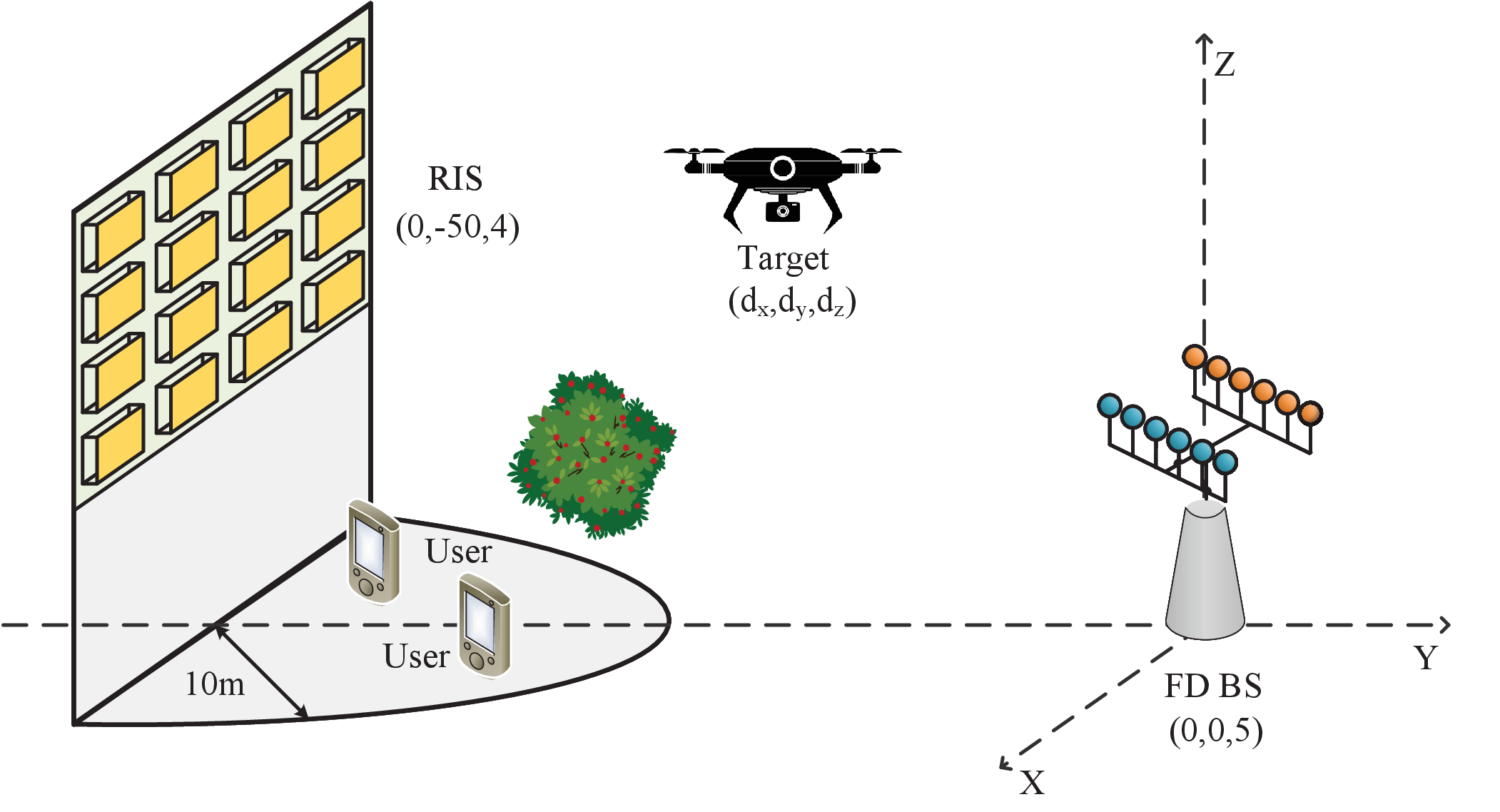}
	\caption{The experiment scenario model.}
	\label{fig.2}
\end{figure}
In this section,
numerical results  are presented to verify our proposals.
The setting of the experiment is shown in Fig. \ref{fig.2},
where one FD BS tries to detect one target and simultaneously serves 4 users covered by a nearby RIS device.
In the experiment,
the BS and the RIS  are located at the three dimensional (3D) coordinates
(0,0,5m)
and
(0,-50m,4m),
respectively.
 The target is randomly distributed at  ($d_x$m,$d_y$m,$d_z$m),
where $d_x \in [-1,1]$,$d_y\in [10,40]$ and $d_z\in [7,10]$.
All users are randomly distributed within a right half circle of the radius of 10m centered at the RIS at an altitude of 1.5m.
The large-scale fading is modeled as ${PL = C_0(d/d_0)}^{-\alpha}$,
where $C_0$ represents the path loss of the reference distance $d_0=1$m,
$d$ and $\alpha$ denote the propagation distance and  the fading exponent, respectively.
{The BS TX-RIS/BS RX-RIS links and self-interference link follow Rician distribution with  Rician factor of 3dB and 5dB, respectively.
The BS TX-user links  and RIS-user links all are assumed to be independent and identically distributed Rayleigh fading channels.
The BS TX-target, BS RX-target and RIS-target links are modeled as line-of-sight (LoS) channels.
The path loss  exponents of
BS TX-User,
BS TX-RIS,
BS RX-RIS,
RIS-User,
RIS-target,
BS TX-target
and
BS RX-target
are
$\alpha_{BU} = 3.6$,
$\alpha_{BTR}= \alpha_{BRR} = 2.7$,
$\alpha_{RU} = 2.4$,
$\alpha_{RT} = 2.2$
and
$\alpha_{BT} = \alpha_{TB} = 2.2$,
respectively.}
The path loss of both self-interference channels is $\rho_{SI} = -110$dB due to the self-interference cancellation \cite{ref37}.
In addition,
the numbers of BS transmit antennas $N_t$ vary from $2$ to $14$ and the number of RIS elements $M$ ranges from $50$ to $1000$.
In most tests, $N_t=N_r=4$ and $M=100$ if not specially stated.
The transmit power for the BS  is set as 30dBm.
The noise power and the predefined target detection level of  BS
are set as $\sigma_{BS}^2  = -90$dBm and $\Gamma_r = 5$dB, respectively.
The RCS is $\sigma_{t}^2 = 1$.

\begin{figure}[t]
	\centering
	\includegraphics[width=.5\textwidth]{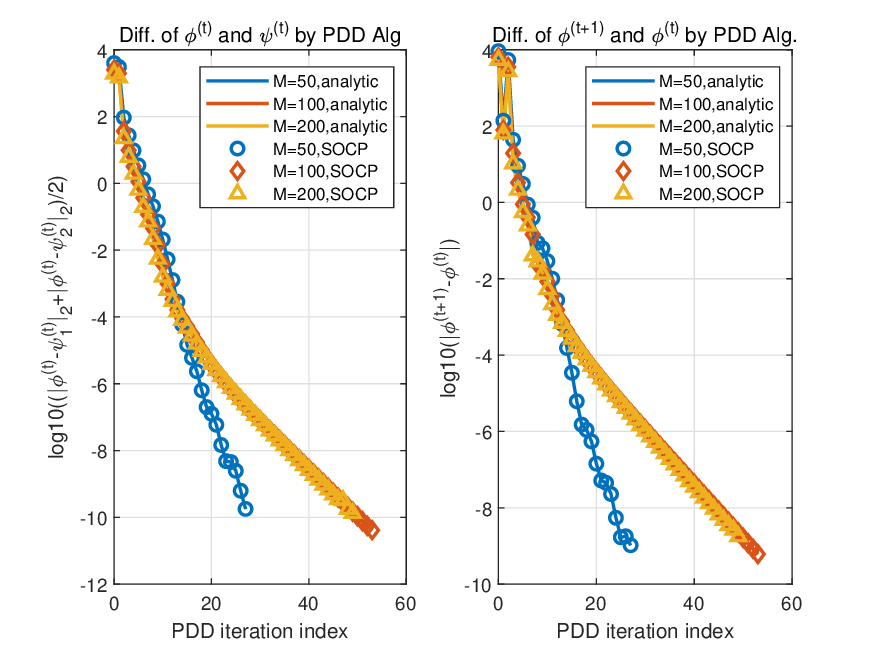}
	\caption{Convergence of PDD method optimizing $\bm{\phi}$ (Alg.1).}
	\label{fig.3}
\end{figure}

\begin{table}[t]
\begin{small}
\centering
\caption{MATLAB Run Time to Update $\bm{\phi}$}
\begin{tabular}{|c|c|c|c|c|c|} \hline \label{PDD_SOCP_vs_KKT}
Method        & $M$=50 & $M$=100 & $M$=200 & $M$=500 & $M$=1000  \\ \hline
SOCP          & 0.3689 & 0.3864  & 0.4910  & 0.5066   & 0.8145  \\ \hline
analytic      & 0.0001 & 0.0004  & 0.0014  & 0.0082   & 0.0444   \\ \hline
\end{tabular}

\end{small}
\end{table}

\begin{table}[t]
\begin{small}
\centering
\caption{MATLAB Run Time to Update $\bm{\psi}_1$}
\begin{tabular}{|c|c|c|c|c|c|} \hline \label{PDD_SOCP_vs_KKT_Psi_1}
Method        & $M$=50 & $M$=100 & $M$=200 & $M$=500 & $M$=1000  \\ \hline
SOCP          & 0.3144 & 0.3242  & 0.3321  & 0.3515   & 0.3684  \\ \hline
analytic      & 0.0001 & 0.0004  & 0.0013  & 0.0081   & 0.0447   \\ \hline
\end{tabular}

\end{small}
\end{table}

Firstly,
Fig. \ref{fig.3}
illustrates the converge behavious of our proposed SOCP-based and analytic-based PDD methods updating the RIS phase-shifts $\bm{\phi}$.
For fair comparison, both the SOCP and KKT implementations start from one common initial point.
In Fig. \ref{fig.3},
under  various settings of number of RIS elements $M$,
the left and right subfigure demonstrate the difference
$\vert \bm{\phi}^{(t)}-\bm{\psi}_{\imath}^{(t)}\vert_2$ and $\vert\bm{\phi}^{(t+1)}-\bm{\phi}^{(t)}\vert_2$
in log domain, respectively, along with the progress of PDD iterations.
As reflected by Fig. \ref{fig.3},
both the SOCP-based and the analytic-based yields nearly identical performance.
The PDD method generally converges well within  $60$ iterations,
i.e., the discrepancy between $\bm{\phi}$ and $\bm{\psi}_{\imath}$ the variation in $\bm{\phi}$ itself is below $10^{-6}$.
However,
it is worth noting that the analytic-based solution has much lower complexity than the SOCP counterpart
(see the following comments for Table \ref{PDD_SOCP_vs_KKT}$\&$\ref{PDD_SOCP_vs_KKT_Psi_1}).

Next,
in Table \ref{PDD_SOCP_vs_KKT}$\&$\ref{PDD_SOCP_vs_KKT_Psi_1},
we examine the complexity of our proposed analytic-based PDD solution.
To this end, we compare the MATLAB runtime of the SOCP-based and the analytic-based methods under different settings of $M$.
Recall that these two competing methods have identical performance as demonstrated in Fig. \ref{fig.3}.
As shown in Table \ref{PDD_SOCP_vs_KKT}$\&$\ref{PDD_SOCP_vs_KKT_Psi_1},
our proposed analytic solutions are highly efficient.
In fact,
the run time of the analytic-based solution is generally one or two orders of magnitude smaller than that of the SOCP one.

\begin{figure}[t]
	\centering
	\includegraphics[width=.5\textwidth]{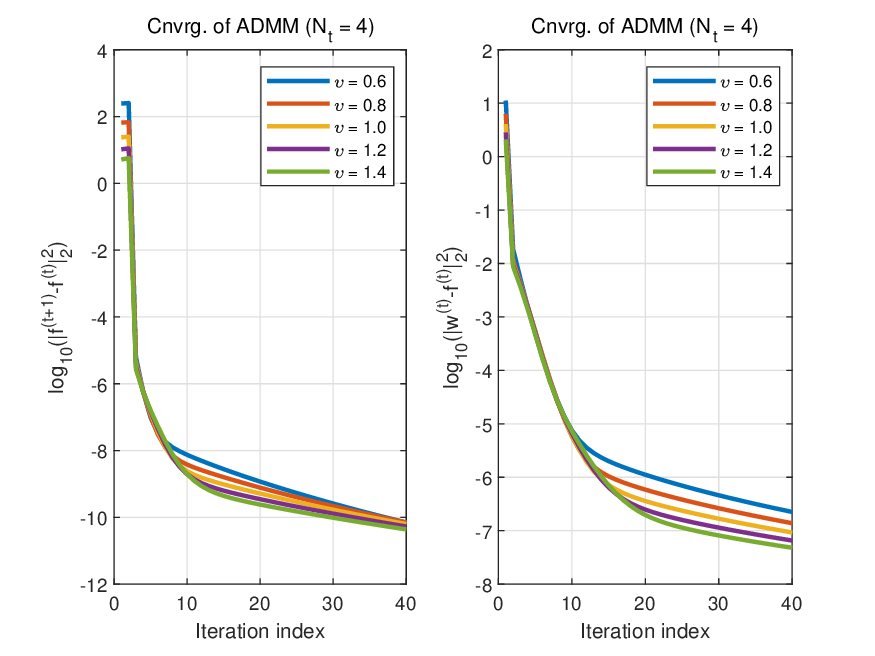}
	\caption{Convergence of ADMM method.}
	\label{fig.4}
\end{figure}

\begin{figure}[t]
	\centering
	\includegraphics[width=.5\textwidth]{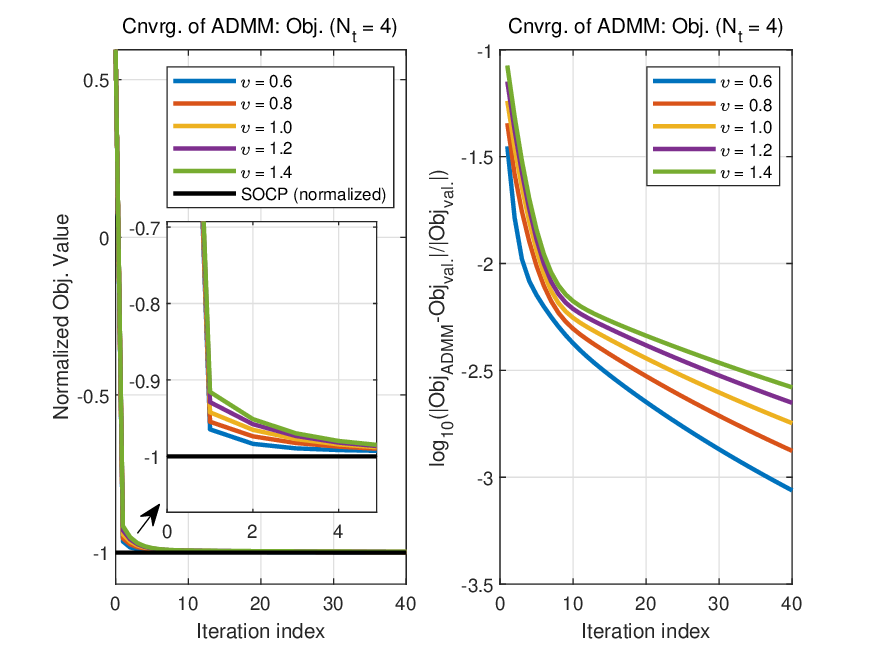}
	\caption{Convergence of objective value of ADMM method.}
	\label{fig.5}
\end{figure}

\begin{table}[t]
\begin{small}
\centering
\caption{MATLAB Run Time to Update $\mathbf{W}$}
\begin{tabular}{|c|c|c|c|c|c|} \hline \label{Table_SOCP_vs_ADMM}
Alg. & $N_t$=2  & $N_t$=4 & $N_t$=8 & $N_t$=12 & $N_t$=14    \\ \hline
SOCP & 0.3139   & 0.3173  & 0.3324  & 0.3632   & 0.4319      \\ \hline
ADMM & 0.0034   & 0.0053  &	0.0319  & 0.1152   & 0.1894      \\ \hline
\end{tabular}

\end{small}
\end{table}

In Fig. \ref{fig.4} and Fig. \ref{fig.5},
we examine the convergence of the analytic-based solution Alg. \ref{alg:ADMM} to optimize $\mathbf{W}$ using ADMM framework.
Note when solving (P24) and (P25),
we rescale their objective and constraint to set the maximal eigenvalue of quadratic coefficient as $1$.
Based on that,
different values of coefficient $\nu$ are tested.
The left and right half of Fig. \ref{fig.4} represents the value of
$\vert \bm{f}^{(t+1)}-\bm{f}^{(t)}\vert_2^2$ and $\vert \bm{f}^{(t)}-\bm{w}^{(t)}\vert_2^2$
in log domain, respectively.
According to Fig. \ref{fig.4}, an appropriate value of the penalty coefficient $\nu$ can be chosen in the range of [0.6,1.4],
which yields sufficient convergence (e.g., precision of $10^{-6}$) within several tens of iterations.

Fig. \ref{fig.5} examines the objective convergence yielded by Alg. \ref{alg:ADMM}.
The left subfigure presents the objective value iterates.
The true objective value of (P11) obtained via utilizing SOCP solver (i.e., CVX) is also presented as a benchmark, which is normalized.
The right half of Fig. \ref{fig.5} represents the difference between the true objective value and that yielded by Alg. \ref{alg:ADMM} in log domain.
Generally, Alg. \ref{alg:ADMM} yields sufficiently accurate objective value within $10$ iterations.

In Table \ref{Table_SOCP_vs_ADMM},
we examine the complexity of the analytic-based solution Alg. \ref{alg:ADMM}.
Under different settings of  the BS antenna numbers $N_t$, the MATLAB runtime of CVX and Alg. \ref{alg:ADMM} are presented in the table.
As shown by the results,
the Alg. \ref{alg:ADMM}'s runtime is smaller than one or two orders of magnitude  of that of the SOCP solver.

\begin{figure}[t]
	\centering
	\includegraphics[width=.5\textwidth]{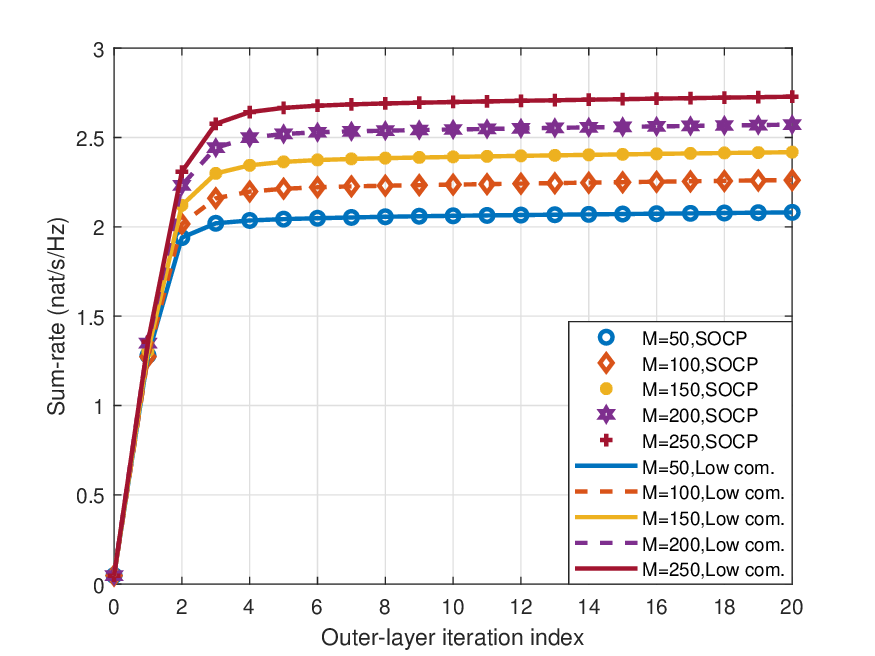}
	\caption{Convergence of Alg. \ref{alg:Overall}.}
	\label{fig.6}
\end{figure}

In Fig. \ref{fig.6},
we check the overall convergence behaviours of our proposed algorithms to tackle the original problem (P1),
including both the SOCP-based and low-complexity (Low com.) solutions.
In all tests, the SOCP-based and analytic-based solutions both start from identical initial points.
As seen from the figure, both algorithms exhibit identical performance and generally converge within $10$ iterations.

\begin{figure}[t]
	\centering
	\includegraphics[width=.5\textwidth]{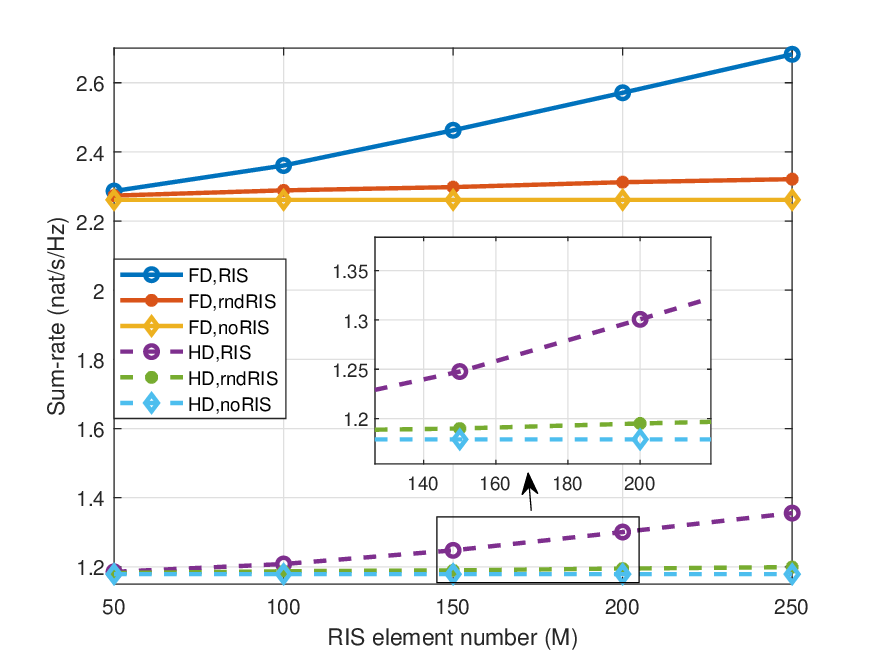}
	\caption{The impact of RIS on communication rate.}
	\label{fig.7}
\end{figure}

In Fig. \ref{fig.7},
we illustrate the sum-rate versus the number of RIS units.
For comparison,
we  consider the  without RIS (``noRIS") and random phase-shift RIS (``rndRIS") schemes
and the  HD system.
It is clearly observe that by increasing the number of RIS' units, the schemes assisted by RIS all monotonically increase the sum-rate.
Moreover,
our proposed algorithm significant outperforms both ``noRIS" and  ``rndRIS" schemes in FD and HD systems,
respectively.
Besides, we can see that the sum-rate of the FD system is larger than that of the
HD system in all schemes.

\begin{figure}[t]
	\centering
	\includegraphics[width=.5\textwidth]{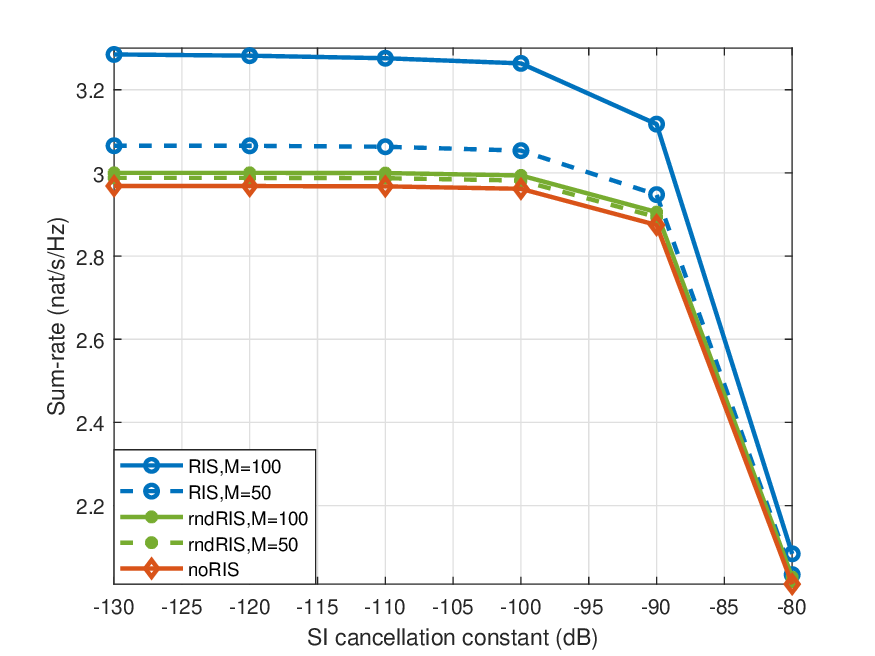}
	\caption{The impact of SI channel.}
	\label{fig.8}
\end{figure}

In Fig.  \ref{fig.8},
we examine the impact of the magnitude of SI channel.
The horizontal axis represents the SI coefficient $\rho_{SI}$,
which is proportional to the magnitude of SI channel $\mathbf{H}_{s}$ \cite{ref37}.
As can be seen,
sum-rate drops when SI increases.
Compared to the no-RIS scenario, the deployment of RIS significantly boosts the sum-rate.

\begin{figure}[t]
	\centering
	\includegraphics[width=.5\textwidth]{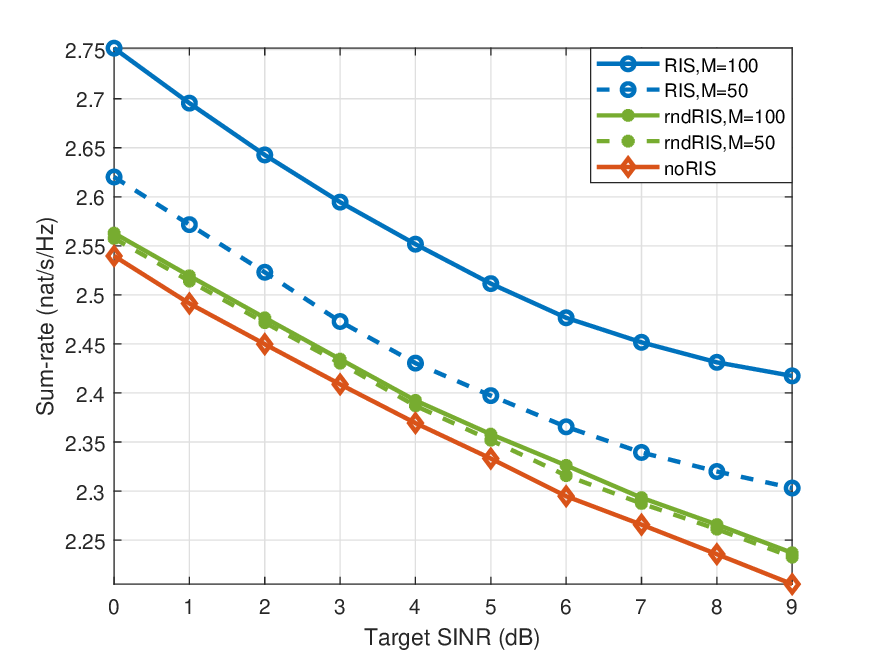}
	\caption{Sum-rate  versus $\Gamma_{r}$.}
	\label{fig.9}
\end{figure}

Fig. \ref{fig.9} demonstrates the achievable sum-rate  versus the predefined target detection level $\Gamma_r$.
It is observed that the sum-rate of all users decreases as $\Gamma_r$ increases,
which reveals the trade-off between the performance of communication and radar sensing.
In addition,
the $M=100$ case can achieve better performance than the $M=50$ case.

\section{Conclusions}
This paper investigates the joint active and passive beamforming design problem in an RIS aided FD ISAC system that performs target sensing and UL communication functionality simultaneously.
We propose an iterative solution to jointly design RIS configuration, users' power allocation
and receiving processors to achieve both radar sensing and communication functionals.
Besides,
we further develop a fully analytic based solution,
which does not depend on any numerical solvers and has low complexity.
Numerical results demonstrate the efficiency and effectiveness of our proposed algorithms and manifest
the benefit of deploying RIS in the uplink FD ISAC system.

\appendix
\subsection{Proof of Lemma 1}
\normalem
Proof:
According to the hypothesis,
strongly duality holds for the problem ($\textrm{P}_{Lm1}$) and it has unique solution since its objective is strictly convex.
We identify ($\textrm{P}_{Lm1}$)'s optimal solution via checking its Karush-Kuhn-Tucker (KKT) conditions \cite{ref38}.
Denote the Lagrangian multiplier associated with the constraint of
($\textrm{P}_{Lm1}$) as $\varsigma$.
Then the KKT conditions of ($\textrm{P}_{Lm1}$) are given as
\begin{subequations} \label{Proof_Lemmma1_KKT}
\begin{align}
& \varsigma \geq 0,
\mathbf{x}^H\mathbf{\bar{Q}}\mathbf{x}-2\textrm{Re}\{\mathbf{\bar{q}}^H\mathbf{x}\}+\bar{q}\leq 0,\label{Proof_Lemma1_0}\\
& \varsigma ( \mathbf{x}^H\mathbf{\bar{Q}}\mathbf{x}-2\textrm{Re}\{\mathbf{\bar{q}}^H\mathbf{x}\}+\bar{q} ) = 0,\label{Proof_Lemma1_1}\\
& \mathbf{{Q}}\mathbf{x}-\mathbf{{q}}
+\varsigma ( \mathbf{\bar{Q}}\mathbf{x}-\mathbf{\bar{q}} ) = 0.\label{Proof_Lemma1_2}
\end{align}
\end{subequations}

Since the $\varsigma$ is non-negative,
we consider two possible cases according to the sign of $\varsigma$ as follows

CASE-I:
if $\varsigma^{\star}=0$,
by  (\ref{Proof_Lemma1_2}),
we can directly obtain
\begin{align}
\mathbf{x}^{\star}=\mathbf{{Q}}^{-1}\mathbf{{q}}.\label{Proof_Lemma1_3}
\end{align}

At this point,
the KKT conditions (\ref{Proof_Lemmma1_KKT}) will be satisfied if and only if the inequality
$(\mathbf{x}^{\star})^{H}\mathbf{\bar{Q}}\mathbf{x}^{\star}-2\textrm{Re}\{\mathbf{\bar{q}}^H\mathbf{x}^{\star}\}+\bar{q}\leq 0$ is satisfied.
If this inequality stands true,
then $\mathbf{x}^{\star}$ in (\ref{Proof_Lemma1_3}) is indeed the optimal solution.
Otherwise, (\ref{Proof_Lemmma1_KKT}) cannot be satisfied and it implies
that $\varsigma^{\star}$ should be positive, as discussed in the following case.

CASE-II:
When $\varsigma^{\star}>0$, by (\ref{Proof_Lemma1_1}) and (\ref{Proof_Lemma1_2})
we have
\begin{align}
\mathbf{x}^{\star} = (\varsigma^{\star}\mathbf{\bar{Q}}+\mathbf{Q})^{-1}(\varsigma^{\star}\mathbf{\bar{q}}+\mathbf{q}).\label{Proof_Lemma1_4}
\end{align}

Since $(\mathbf{{Q}}^{-1}\mathbf{{q}})^{ H}\mathbf{\bar{Q}}(\mathbf{{Q}}^{-1}\mathbf{{q}})
-2\textrm{Re}\{\mathbf{\bar{q}}^H(\mathbf{{Q}}^{-1}\mathbf{{q}})\}+\bar{q}> 0$,
hence there have a unique positive $\varsigma^{\star}$ satisfying
\begin{align}
&\big((\varsigma^{\star}\mathbf{\bar{Q}}+\mathbf{Q})^{-1}(\varsigma^{\star}\mathbf{\bar{q}}+\mathbf{q})\big)^H
\mathbf{\bar{Q}}
(\varsigma^{\star}\mathbf{\bar{Q}}+\mathbf{Q})^{-1}(\varsigma^{\star}\mathbf{\bar{q}}+\mathbf{q})\nonumber\\
&-2\textrm{Re}\{\mathbf{\bar{q}}^H(\varsigma^{\star}\mathbf{\bar{Q}}+\mathbf{Q})^{-1}(\mathbf{\varsigma^{\star}\bar{q}}+\mathbf{q})\}+\bar{q}= 0.
\end{align}
The unique $\varsigma^{\star}$ can be obtained by the  Newton's method.
Therefore, Lemma \ref{lem:complex_Convex_trust_region problem} has been proved.

\subsection{Proof of Theorem 1}
\normalem
Proof:
Firstly,
one critical observation is that we can just assume that
$b_k<0$
for all
$k\in\mathcal{K}$
without loss of optimality.
In fact,
if $b_k\geq0$ for some specific $k$,
then obviously
$p_k^{\star}=0$
minimizes the summand associated with
$p_k$
in the objective while not affecting the power allocation to other users as constrained in (\ref{Q_sun_power_constraint}).

Next, we consider to determine the optimal solutions of (P26) when $b_k<0$, $\forall k \in \mathcal{K}$.
Obviously,
Slater's condition for (P26) when all $p_k$'s take sufficiently small positive values.
Therefore,
strong duality holds for (P26) and we determine its optimal value via analyzing its KKT conditions.

Via introducing the Lagrangian multipliers  $\{\eta_k\}$, $\{\gamma_k\}$ and $\nu$
associated with the constraints of (P26), $0 \leq p_k$, $p_k \leq \bar{{P}}_{U,k}$ and ${\sum}_{k=1}^{K}d_k p_k^2 \leq \hat{c}_5$,
respectively,
the KKT conditions  are given as
\begin{small}
\begin{subequations}
\begin{align}
&0 \leq p_k\leq \bar{P}_{U,k},
\gamma_k \geq 0,
\eta_k \geq 0,
\nu \geq 0,
{\sum}_{k=1}^{K}\!\!d_k p_k^2\! -\! \hat{c}_5 \!\leq\! 0,\label{Proof of Theorem 1_1}\\
&\gamma_k(p_k- \bar{P}_{U,k}) = 0,
\eta_k p_k = 0,
\nu({\sum}_{k=1}^{K}d_k p_k^2 - \hat{c}_5) = 0,\label{Proof of Theorem 1_2}\\
& 2 a_k p_k + b_k + \gamma_k - \eta_k + 2\nu d_k p_k = 0,\
\forall k \in \mathcal{K}.\label{Proof of Theorem 1_3}
\end{align}
\end{subequations}
\end{small}

In the following,
we will analyze the KKT conditions
in two possible cases according to the sign value of $\nu$,
i.e., $\nu=0$ or $\nu > 0$.

\uline{CASE-I}:
$\nu^{\star}=0$.
Via (\ref{Proof of Theorem 1_1}),
we can have
${\sum}_{k=1}^{K}d_k p_k^{\star 2} \leq \hat{c}_5$.
Then we investigate the value of $\gamma_k$.

\uwave{case-i}:
if $\gamma_k^{\star}>0$, by (\ref{Proof of Theorem 1_2}),
we immediately have
$p_k^{\star} = \bar{P}_{U,k}$ and hence $\eta_k^{\star}=0$.
Therefore,
(\ref{Proof of Theorem 1_3}) is equivalently rewritten as
\begin{align}
\gamma_k^{\star} = -(2 a_k \bar{P}_{U,k} + b_k) > 0.
\end{align}
Since $\gamma_k^{\star}>0$,
 $\bar{P}_{U,k}$ need to satisfy
$\bar{P}_{U,k} < (-b_k)/(2 a_k) $.
Otherwise,
$\gamma_k^{\star}$ will not be positive and then this sub-case can not occur.

\uwave{case-ii}:
if $\gamma_k^{\star}=0$,
via  (\ref{Proof of Theorem 1_2}),
we have $0 \leq p_k^{\star} \leq \bar{P}_{U,k}$ and $\eta_k^{\star}\geq0$.
The equation (\ref{Proof of Theorem 1_3}) is transformed into
\begin{align}
& \gamma_k^{\star} = -2 a_k p_k^{\star} - b_k + \eta_k^{\star} =0,\ \forall k \in \mathcal{K}.\label{Proof_theorem_1_gamma}
\end{align}
Then,
we will obtain
\begin{align}
&p_k^{\star} = ({\eta_k^{\star}-b_k})/({2a_k}),\\
&\eta_k^{\star}=2 a_k p_k^{\star} + b_k.\label{eta_65}
\end{align}
Since $p_k^{\star}$ lies in the range  $[0,\bar{P}_{U,k}]$,
$\eta_k^{\star}$ also has a bounded range $[b_k,2 a_k \bar{P}_{U,k} + b_k]$.
Since
$\eta_k^{\star}\geq0$,
this is possible only if the upper-bound $2 a_k \bar{P}_{U,k} + b_k \geq 0$,
i.e., $({-b_k})/({2a_k})\leq  \bar{P}_{U,k}$.
Otherwise, \uwave{case-ii} could not occur.

Besides,
we notice that $\eta_k^{\star}=0$.
In fact, if $\eta_k^{\star}>0$, then $ p_k^{\star}=0$ and hence (\ref{eta_65}) implies $\eta_k^{\star}=b_k$.
Since $b_k<0$ as previously discussed,
the equality $\eta_k^{\star}=b_k$ cannot stand.
Therefore,
$\eta_k^{\star}=0$ and the optimal solution $p_k^{\star} = ({-b_k})/({2a_k})$.

Summarizing the above two sub-cases,
we readily obtain
\begin{subequations}
\begin{align}
&\gamma_k^{\star}  = [-(2 a_k \bar{P}_{U,k} + b_k)]^{+},\\
&p_k^{\star} = \mathrm{min}\{\bar{P}_{U,k},(-{b_k})/({2 a_k})\},
\forall k \in \mathcal{K},\label{Proof_Theorem_1_nu_0}
\end{align}
\end{subequations}
where $[x]^{+} \triangleq \mathrm{max}\{x,0\}$.
Note that (\ref{Proof_Theorem_1_nu_0}) will satisfy all KKT conditions in (\ref{Proof of Theorem 1_1})-(\ref{Proof of Theorem 1_3}) except for the sum-power constraint
${\sum}_{k=1}^{K}d_k p_k^{\star 2} \leq \hat{c}_5$.
If this constraint is satisfied,
the value of $\{p_k\}$ in (\ref{Proof_Theorem_1_nu_0}) is the optimal solution to problem (P26).
However,
if ${\sum}_{k=1}^{K}d_k p_k^{\star 2} > \hat{c}_5$,
\uline{CASE-I} would not occur,
and then we further to consider \uline{CASE-II},
i.e., $\nu>0$.

\uline{CASE-II}: $\nu^{\star}>0$.
By (\ref{Proof of Theorem 1_2}),
we have ${\sum}_{k=1}^{K}d_k p_k^{\star 2} = \hat{c}_5$.

Similar to  \uline{CASE-I},
we also consider two possible subcases of $\gamma_k$ in the following.

\uwave{case-i}:
If $ \gamma_k^{\star} > 0$,
then by (\ref{Proof of Theorem 1_2}),
we have
$p_k^{\star} = \bar{P}_{U,k}$ and $\eta_k^{\star}=0$.
The equation (\ref{Proof of Theorem 1_3}) is reduced to
\begin{align}
\gamma_k^{\star}  + 2 a_k \bar{P}_{U,k} + b_k + 2 \nu^{\star} d_k \bar{P}_{U,k}=0,\label{Proof_of_Theorom_CASEII_casei_gamma}
\end{align}
which is equivalent to
\begin{align}
\gamma_k^{\star}  = -2 a_k \bar{P}_{U,k} - b_k   - 2 \nu^{\star} d_k \bar{P}_{U,k}.
\end{align}
Since $\gamma_k^{\star}>0$ by assumption,
this requires
\begin{align}
\nu^{\star} <  -\frac{a_k}{d_k}-\frac{b_k}{2 d_k \bar{P}_{U,k}}.
\end{align}
Therefore,
associated with $\nu^{\star}>0$,
the following condition
\begin{align}
0 <\nu^{\star} <  -\frac{a_k}{d_k}-\frac{b_k}{2 d_k \bar{P}_{U,k}}\label{Proof_theorem_CASE_II_nu_1}
\end{align}
should be satisfied.
Otherwise,
\uwave{case-i} could not occur.

\uwave{case-ii}:
If $\gamma_k^{\star} = 0$,
via (\ref{Proof of Theorem 1_2})
we have
$0 < p_k^{\star} \leq \bar{P}_{U,k}$ and $\eta_k^{\star}\geq0$.
Therefore, by (\ref{Proof of Theorem 1_3}),
we obtain
\begin{align}
&p_k^{\star} = \frac{\eta_k^{\star}-b_k}{2a_k+2 \nu^{\star} d_k},\\
&\eta_k^{\star} =  2 a_k p_k^{\star} + b_k   + 2 \nu^{\star} d_k p_k^{\star}.
\end{align}
Since $p_k^{\star}$ has a range of $[0,\bar{P}_{U,k}]$,
$\eta_k^{\star}$ takes value in the range $[b_k, 2 (a_k+\nu^{\star} d_k) \bar{P}_{U,k} + b_k ]$.
Due to $\eta_k^{\star}\geq0$,
the upper-bound of $\eta_k^{\star}$ should satisfy $2 (a_k+\nu^{\star} d_k) \bar{P}_{U,k} + b_k\geq0$.
Therefore,
the $\nu^{\star}$ can only take value in the range as follows
\begin{align}
\nu^{\star}\geq-\frac{a_k}{d_k}-\frac{b_k}{2 d_k \bar{P}_{U,k}}.\label{Proof_theorem_CASE_II_nu_2}
\end{align}

Following similar arguments as in \underline{CASE-I},
it can be shown that $\eta_k^{\star}=0$.
Therefore,
 $\eta_k^{\star}=0$ holds
and  the optimal solution $p_k^{\star}$ is given as
\begin{align}
p_k^{\star} = -\frac{b_k}{2a_k+2 \nu^{\star} d_k}.
\end{align}

If $\nu^{\star}<-\frac{a_k}{d_k}-\frac{b_k}{2 d_k \bar{P}_{U,k}}$,
$\eta_k^{\star}$ will be negative and
this sub-case could not occur indeed.
Summarizing the above two sub-cases and comparing the conditions in
(\ref{Proof_theorem_CASE_II_nu_1})
and
(\ref{Proof_theorem_CASE_II_nu_2}),
we have the optimal solution as follows
\begin{align}
p_k^{\star}(\nu^{\star}) = \bigg[-\frac{b_k}{2a_k + 2 \nu^{\star} d_k }\bigg]^{\bar{P}_{U,k}}.\label{Proof_theorem_1_p_solution_2}
\end{align}
Note that $p_k^{\star}(\nu^{\star})$ is a monotonically decreasing function in $\nu^{\star}$
and $p_k^{\star}(0) = \textrm{min}\{-\frac{b_k}{2a_k} ,\bar{P}_{U,k}\}=\hat{p}_k^{\star}$.
Therefore,
when $\nu^{\star}$ decreases $+\infty$ to $0$,
the value of $p_k^{\star}(\nu^{\star})$ varies $0$ to $\hat{p}_k^{\star}$.
According to ${\sum}_{k=1}^{K}d_k \hat{p}_k^{\star 2} > \hat{c}_5$ in \uline{CASE-II},
hence there exists an unique positive $\nu^{\star}$ satisfying ${\sum}_{k=1}^{K}d_k ({p}_k^{\star}(\nu^{\star}))^2 = \hat{c}_5$.

\subsection{Proof of Lemma 3 }
\normalem
Proof:
Consider the following optimization problem
\begin{subequations}
\begin{align}
\textrm{(P27)}:\mathop{\textrm{min}}
\limits_{
\{p_k\}}\
&   {\sum}_{k=1}^{K}a_k p_k^2 + {\sum}_{k=1}^{K}b_k p_k\\
\textrm{s.t.}\
& 0 \leq p_k\leq \bar{P}_{U,k}, \forall k \in \mathcal{K},
\end{align}
\end{subequations}
where the problem (P27) is a relaxation of (P26) by ignoring the sun-power constraint (\ref{Q_sun_power_constraint}).

According to the definition of $\{\hat{p}_k\}$ given in Theorem \ref{alg:p_KKT}, it is easily to prove that $\{\hat{p}_k\}$ is indeed the optimal solution of (P27).
Hence, we have $\mathsf{obj}(\{{p}_k^{\star}\})\geq \mathsf{obj}(\{\hat{p}_k\})$.

Next, the  problem (P26) is rewritten in  an equivalent form as follows:
\begin{subequations}
\begin{align}
\textrm{(P28)}:\mathop{\textrm{min}}
\limits_{
\mathbf{Q}}\
&   {\sum}_{k=1}^{K}a_k p_k^2 + {\sum}_{k=1}^{K}b_k p_k\\
\textrm{s.t.}\
& {\sum}_{k=1}^{K}d_k p_k^2 \leq \hat{c}_4,
\end{align}
\end{subequations}
where we define the convex set $\mathcal{Q}$ as
$\mathcal{Q} \triangleq \{ \{p_k\} \vert 0 \leq p_k \leq \bar{P}_{U,k}, \forall k \in \mathcal{K} \}$.
The Lagrangian of (P28) is formulated as
\begin{small}
\begin{align}
\mathcal{L}(\{p_k\},\nu)\! =\!\! {\sum}_{k=1}^{K}\!a_k p_k^2\! +\!\! {\sum}_{k=1}^{K}\!b_k p_k \!+\! \nu({\sum}_{k=1}^{K}\!d_k p_k^2\! -\!\hat{c}_4).
\end{align}
\end{small}
\vspace{-0.5cm}

\noindent
Since the Slater's condition of problem (P28) is obviously satisfied,
and then strong duality holds.
Therefore, we can obtain a saddle point of the Lagrangian in (P28) \cite{ref38}
 via
a pair of optimal primal-dual variables $(\{p_k^{\star}\},\nu^{\star})$.
Then we have the following relations
\begin{align}
&\mathsf{obj}(\{\hat{p}_k\})\leq\mathsf{obj}(\{{p}_k^{\star}\}) = \mathcal{L}(\{p_k^{\star}\},\nu^{\star})\label{Proof_Lemma_3}\\
&\overset{(a)}=\mathop{\textrm{min}}
\limits_{\{p_k\}\in \mathbf{Q}}
\big\{  \! {\sum}_{k=1}^{K}\!a_k p_k^2\! +\!\! {\sum}_{k=1}^{K}\!b_k p_k \!+\! \nu^{\star}({\sum}_{k=1}^{K}\!d_k p_k^2\! -\!\hat{c}_5)  \big\}\nonumber\\
&\overset{(b)}\leq  {\sum}_{k=1}^{K}\!a_k \tilde{p}_k^2\! +\!\! {\sum}_{k=1}^{K}\!b_k \tilde{p}_k \!+\! \nu^{\star}({\sum}_{k=1}^{K}\!d_k \tilde{p}_k^2\! -\!\hat{c}_5),\nonumber
\end{align}
where $(a)$ is due to the saddle point theorem \cite{ref38}
and $(b)$ holds because $\{\tilde{p}_k\}$ ia an arbitrary strictly feasible solution to problem (P26).
Note that $ {\sum}_{k=1}^{K}d_k \tilde{p}_k^2 < \hat{c}_5 $ is satisfied according to the chose of $\{\tilde{p}_k\}$.
Lastly, we rearrange the equality (\ref{Proof_Lemma_3}) and then obtain the upper-bound in (\ref{Lemma3_upper_bound}).


\end{document}